\documentclass[preprint,prd,showpacs,showkeys,eqsecnum,superscriptaddress,nofootinbib]{revtex4}
\usepackage{epsfig}

\usepackage{amsmath, amssymb, amsfonts, graphics}
\date{\today}
 \normalsize

\newcommand{\bmat}{\left(\begin{array}}
\newcommand{\emat}{\end{array}\right)}
\newcommand{\be}{\begin{equation}}
\newcommand{\ee}{\end{equation}}
\newcommand{\bea}{\begin{eqnarray}}
\newcommand{\eea}{\end{eqnarray}}

\def\gtwid{\mathrel{\raise.3ex\hbox{$>$\kern-.75em\lower1ex\hbox{$\sim$}}}}
\def\ltwid{\mathrel{\raise.3ex\hbox{$<$\kern-.75em\lower1ex\hbox{$\sim$}}}}
\def\gev{{\rm \, Ge\kern-0.125em V}}
\def\tev{{\rm \, Te\kern-0.125em V}}

\def    \be            {\begin{equation}}
\def    \ee            {\end{equation}}
\def    \bea           {\begin{eqnarray}}
\def    \eea           {\end{eqnarray}}

 \normalsize

\def\gtwid{\mathrel{\raise.3ex\hbox{$>$\kern-.75em\lower1ex\hbox{$\sim$}}}}
\def\ltwid{\mathrel{\raise.3ex\hbox{$<$\kern-.75em\lower1ex\hbox{$\sim$}}}}
\def\gev{{\rm \, Ge\kern-0.125em V}}
\def\tev{{\rm \, Te\kern-0.125em V}}

\def    \be            {\begin{equation}}
\def    \ee            {\end{equation}}
\def    \bea           {\begin{eqnarray}}
\def    \eea           {\end{eqnarray}}

\begin{document}
\renewcommand{\thefootnote}{\fnsymbol{footnote}}
\vspace{.3cm}
%

\title{\Large\bf A Two-Singlet Model for Light Cold Dark Matter }

\author{{\bf Abdessamad Abada}}
\email{a.abada@uaeu.ac.ae}
\affiliation{Laboratoire de Physique des Particules\ et Physique Statistique, Ecole
Normale Sup\'{e}rieure, BP 92 Vieux Kouba, 16050 Alger, Algeria.}
\altaffiliation{Present address: Physics Department, UAE University, POB 17551, Al Ain, UAE .}
\author{{\bf Salah Nasri}}
\email{snasri@uaeu.ac.ae}
\affiliation{Physics Department, UAE University , Al Ain, United Arab Emirates.}
\author{{\bf Djamal Ghaffor}}
\email{dghaffor@mail-enset.dz}
\affiliation{ Laboratoire de Physique theorique d'oran -ES-SENIA University , 31000 Oran Algeria.}












\vspace{1.2cm}


\vspace{1.5cm}

\begin{abstract}

We extend the Standard Model by adding two gauge-singlet $\mathbb{Z}_{2}$%
-symmetric scalar fields that interact with visible matter only through the
Higgs particle. One is a stable  dark matter WIMP, and the other one
undergoes a spontaneous breaking of the symmetry that opens new channels for
the dark matter annihilation, hence lowering the mass of the WIMP. We study
the effects of the observed   dark matter relic  abundance on the
annihilation cross section and find that in most regions of the parameters space, light
dark matter is viable. We also compare the elastic scattering  cross-section of our dark matter candidate off nucleus 
with existing (CDMSII and XENON100) and projected (SuperCDMS and XENON1T) experimental
exclusion bounds. We find that most of the allowed mass range for light dark matter will be probed by the projected sensitivity of XENON1T experiment.
\end{abstract}

\date{\today }
\maketitle




\section{Introduction}

Cosmology tells us that about 25\% of the total mass density in the
Universe is dark matter that cannot be accounted for by conventional baryons
\cite{Observ}. Alongside observation, intense theoretical efforts
are made in order to elucidate the nature and properties of this unknown
form of matter. In this context, electrically neutral and colorless weakly
interacting massive particles (WIMPs) form an attractive scenario. Their
broad properties are: masses in the range of one to a few hundred GeV,
coupling constants in the milli-weak scale and lifetimes longer than  the age
of the Universe.

Recent data from the direct-detection experiments DAMA/LIBRA \cite{dama-libra} and CoGeNT \cite%
{cogent}, and the recent analysis of the data from the Fermi Gamma Ray Space Telescope \cite{HG},  if interpreted as signal for  dark matter,   require light WIMPs in the range of 5 to 10 GeV   \cite{Groups}.  Also, galactic substructure requires still lighter dark matter masses  \cite{dave-et-al, mcdonald-2}. In this regard, it is useful to note in passing that the XENON100
collaboration has provided serious constraints  on the region of interest  to DAMA/LIBRA and CoGeNT \cite{xenon100}, assuming a constant extrapolation of the liquid xenon scintillation response for nuclear recoils below 5 keV, a claim disputed in \cite{collar-mckinsey}. Most recently,  the CDMS collaboration has released the analysis of their low-energy threshold  data \cite{Ge2010} which seems to exclude the parameter space for dark matter interpretation of DAMA/LIBRA and CoGeNT results, assuming a standard halo dark matter model with an escape velocity $v_{esc} = 544$ km/s and   neglecting the effect of ion channeling \cite{Chann}. However, with a highly anisotropic velocity distribution, it may be possible to reconcile  the CoGeNT and DAMA/LIBRA results with the current exclusion limits from CDMS and XENON \cite{FS} (see also comments on p.6 in \cite{KH} about the possibility of shifting the exclusion contour in \cite{Ge2010} above the CoGeNT signal region). In addition,  CRESST,  another direct detection experiment at Gran Sasso, which uses CaWO$_4$ as target material, reported in  talks at the IDM 2010   and WONDER 2010 workshops an excess   of events in their  oxygen band  instead of tungsten band. If this signal is not due to neutron background a possible interpretation   could be the elastic  scattering of a light WIMP  depositing a detectable recoil energy on the the lightest nuclei (oxygen) in the detector \cite{CRESST}. While this result has to await confirmation from the CRESST collaboration, it is clear that it is  important as well as interesting to study light dark matter.


The most popular candidate for dark matter is the neutralino, a neutral  $R$-odd supersymmetric particle. Indeed, they  are only produced or destroyed in pairs, thus rendering the lightest SUSY particle (LSP) stable \cite{LSP}. In the minimal version of the supersymmetric
extension of the Standard Model, the neutralino is a linear combination of the fermionic  partners of the
neutral electroweak gauge bosons (gauginos) and the neutral Higgs bosons (higgsinos). They can annihilate
through a t-channel sfermion exchange  into standard model fermions,  or via a t-channel  chargino-mediated process into $W^+W^-$, or through an s-channel pseudoscalar Higgs exchange into fermion pairs and they can  undergo elastic scattering with nuclei mainly through scalar Higgs exchange \cite{Kamio}. If these  neutralinos were light,  they would then be
overproduced in the early universe and, in the minimal model, would not have
an elastic-scattering cross section large enough to account for the DAMA/LIBRA and CoGeNT
results due to constraints from other experiments
such as the LEP, Tevatron, and rare decays \cite{ Nathal, Betal, HT}.

Therefore, with no clear clue yet as to what the internal structure of these
WIMPs is, if any, a pedestrian approach can be attractive. In this logic,
the simplest of models is to extend the Standard Model by adding a real
scalar field, the dark matter, a Standard-Model gauge singlet that interacts
with visible particles via the Higgs field only. To ensure stability, it is
endowed with a discrete $\mathbb{Z}_{2}$ symmetry that does not
spontaneously break. Such a model can be seen as a low-energy remnant of
some higher-energy physics waiting to be understood. In this cosmological
setting, such an extension has first been proposed in \cite{silveira-zee}
and further studied in \cite{mcdonald-1} where the unbroken $\mathbb{Z}_{2}$
symmetry is extended to a global U(1) symmetry. A more extensive exploration
of the model and its implications was done in \cite%
{burgess-pospelov-veldhuis}, specific implications on Higgs detection and
LHC physics discussed in \cite{barger et al} and one-loop vacuum stability
looked into and perturbativity bounds obtained in \cite{gonderinger et al}. The work of  \cite{he-li-li-tandean-tsai} considers also this minimal extension and uses constraints from the experiments XENON10 \cite{xenon10-1} and
CDMSII \cite{cdmsII-1} to exclude dark matter masses smaller than 50, 70 and
75GeV for Higgs masses equal to 120, 200 and 350 GeV respectively.

In order to allow for light dark matter, it is therefore necessary to go
beyond the minimal one-real-scalar extension of the Standard Model. The
natural  next step is to add another real scalar field, endowed
with a $\mathbb{Z}_{2}$ symmetry too, but one which is spontaneously broken
so that new channels for dark matter annihilation are opened, increasing
this way the annihilation cross-section, hence allowing smaller masses. This
auxiliary field must also be a Standard Model gauge singlet.

 After this brief introductory motivation, we present the model in the next section. We
perform the spontaneous breaking of the electroweak and the additional $%
\mathbb{Z}_{2}$ symmetries in the usual way. We clarify the physical modes
as well as the physical parameters. There is mixing between the physical new
scalar field and the Higgs, and this is one  of the quantities parametrizing the
subsequent physics. In section three, we impose the constraint from the
known dark matter relic density on the dark-matter annihilation cross
section and study its effects. Of course, as we will see, the parameter
space is quite large, and so, it is not realistic to hope to cover all of it
in one single work of acceptable size. Representative values have to be
selected and the behavior of the model as well as its capabilities are
described. Our main focus in this study is the mass range 0.1GeV -- 100GeV
and we find that the model is rich enough to bear dark matter in most of it,
including the very light sector. In section four, we determine the total
cross section $\sigma _{\det }$ for non relativistic elastic scattering of
dark matter off a nucleon target and compare it to the current direct-detection
experimental bounds and projected sensitivity. For this, we choose the results of
CDMSII  and XENON100 , and the projections of
SuperCDMS \cite{supercdms} and XENON1T \cite{xenon1t}. Here too we cannot
cover all of the parameter space nor are we going to give a detailed account
of the behavior of $\sigma _{\det }$ as a function of the dark matter mass,
but general patterns are mentioned. The last section is devoted to some concluding remarks. Note that as a rule, we have avoided in this first study
narrowing the choice of parameters using particle phenomenology. Of course,
such phenomenological constraints have to be addressed ultimately and this
is left to a forthcoming investigation, contenting ourselves in the present
work with a limited set of remarks mentioned in this last section. Finally,
we have gathered in an appendix the partial results regarding the
calculation of the dark matter annihilation cross section.

\section{A two-singlet model for dark matter}

We extend the Standard Model by adding two real, spinless and $\mathbb{Z}%
_{2} $-symmetric fields: the dark matter field $S_{0}$ for which the $%
\mathbb{Z}_{2}$ symmetry is unbroken and an auxiliary field $%
\chi _{1}$ for which it is spontaneously broken. Both fields are Standard
Model gauge singlets and hence can interact with `visible' particles only
via the Higgs doublet $H$. This latter is taken in the unitary gauge such
that $H^{\dagger }=1/\sqrt{2}\left( 0\hspace{6pt}h^{\prime }\right) $, where
$h^{\prime }$ is a real scalar. We assume all processes calculable in
perturbation theory. The potential function that includes $S_{0}$, $%
h^{\prime }$ and $\chi _{1}$ writes as follows:%
\begin{equation}
U=\frac{\tilde{m}_{0}^{2}}{2}S_{0}^{2}-\frac{\mu ^{2}}{2}h^{\prime 2}-\frac{%
\mu _{1}^{2}}{2}\chi _{1}^{2}+\frac{\eta _{0}}{24}S_{0}^{4}+\frac{\lambda }{%
24}h^{\prime 4}+\frac{\eta _{1}}{24}\chi _{1}^{4}+\frac{\lambda _{0}}{4}%
S_{0}^{2}h^{\prime 2}+\frac{\eta _{01}}{4}S_{0}^{2}\chi _{1}^{2}+\frac{%
\lambda _{1}}{4}h^{\prime 2}\chi _{1}^{2},  \label{U}
\end{equation}%
where $\tilde{m}_{0}^{2}$, $\mu ^{2}$and $\mu _{1}^{2}$ and all the coupling
constants are real positive  numbers. In the Standard Model scenario,
electroweak spontaneous symmetry breaking occurs for the Higgs field, which
then oscillates around the vacuum expectation value $v=246\mathrm{GeV}$ \cite%
{particle-data-group}. The field $\chi _{1}$ will oscillate around the
vacuum expectation value $v_{1}>0$. Both $v$ and $v_{1}$ are related to the
parameters of the theory by the two relations:%
\begin{equation}
v^{2}=6\frac{\mu ^{2}\eta _{1}-6\mu _{1}^{2}\lambda _{1}}{\lambda \eta
_{1}-36\lambda _{1}^{2}};\qquad v_{1}^{2}=6\frac{\mu _{1}^{2}\lambda -6\mu
^{2}\lambda _{1}}{\lambda \eta _{1}-36\lambda _{1}^{2}}.  \label{vacua}
\end{equation}%
It is assumed that the self-coupling constants are sufficiently larger than
the mutual ones.

Writing $h^{\prime }=v+\tilde{h}$ and $\chi _{1}=v_{1}+\tilde{S}_{1}$, the
potential function becomes, up to an irrelevant zero-field energy:%
\begin{equation}
U=U_{\mathrm{quad}}+U_{\mathrm{cub}}+U_{\mathrm{quar}},  \label{U rewritten}
\end{equation}%
where the mass-squared (quadratic) terms are gathered in $U_{\mathrm{quad}}$%
, the cubic interactions in $U_{\mathrm{cub}}$ and the quartic ones in $U_{%
\mathrm{quar}}$. The quadratic terms are given by:%
\begin{equation}
U_{\mathrm{quad}}=\frac{1}{2}m_{0}^{2}S_{0}^{2}+\frac{1}{2}M_{h}^{2}\tilde{h}%
^{2}+\frac{1}{2}M_{1}^{2}\tilde{S}_{1}^{2}+M_{1h}^{2}\tilde{h}\tilde{S}_{1},
\label{U_quad}
\end{equation}%
where the mass-squared coefficients are related to the original parameters
of the theory by the following relations:%
\begin{eqnarray}
m_{0}^{2} &=&\tilde{m}_{0}^{2}+\frac{\lambda _{0}}{2}v^{2}+\frac{\eta _{01}}{%
2}v_{1}^{2};\quad M_{h}^{2}=-\mu ^{2}+\frac{\lambda }{2}v^{2}+\frac{\lambda
_{1}}{2}v_{1}^{2};  \notag \\
M_{1}^{2} &=&-\mu _{1}^{2}+\frac{\lambda _{1}}{2}v^{2}+\frac{\eta _{1}}{2}%
v_{1}^{2};\quad M_{1h}^{2}=\lambda _{1}v\,v_{1}.  \label{masses}
\end{eqnarray}%
Replacing the vacuum expectation values $v$ and $v_{1}$ by their respective
expressions (\ref{vacua}) will not add clarity.
In this field basis, the mass-squared matrix is not diagonal: there is
mixing between the fields $\tilde{h}$ and $\tilde{S}_{1}$. Denoting the
physical mass-squared field eigenmodes by $h$ and $S_{1}$, we rewrite:%
\begin{equation}
U_{\mathrm{quad}}=\frac{1}{2}m_{0}^{2}S_{0}^{2}+\frac{1}{2}m_{h}^{2}h^{2}+%
\frac{1}{2}m_{1}^{2}S_{1}^{2},  \label{U_quad diagonal}
\end{equation}%
where the physical fields are related to the mixed ones by a $2\times 2$
rotation:%
\begin{equation}
\begin{pmatrix}
h \\
S_{1}%
\end{pmatrix}%
=%
\begin{pmatrix}
\cos \theta & \sin \theta \\
-\sin \theta & \cos \theta%
\end{pmatrix}%
\begin{pmatrix}
\tilde{h} \\
\tilde{S}_{1}%
\end{pmatrix}%
.  \label{rotation}
\end{equation}%
Here $\theta $ is the mixing angle, related to the original mass-squared
parameters by the relation:%
\begin{equation}
\tan 2\theta =\frac{2M_{1h}^{2}}{M_{1}^{2}-M_{h}^{2}},  \label{theta}
\end{equation}%
and the physical masses in (\ref{U_quad diagonal}) by the two relations:%
\begin{eqnarray}
m_{h}^{2} &=&\frac{1}{2}\left[ M_{h}^{2}+M_{1}^{2}+\varepsilon \left(
M_{h}^{2}-M_{1}^{2}\right) \sqrt{\left( M_{h}^{2}-M_{1}^{2}\right)
^{2}+4M_{1h}^{4}}\right] ;  \notag \\
m_{1}^{2} &=&\frac{1}{2}\left[ M_{h}^{2}+M_{1}^{2}-\varepsilon \left(
M_{h}^{2}-M_{1}^{2}\right) \sqrt{\left( M_{h}^{2}-M_{1}^{2}\right)
^{2}+4M_{1h}^{4}}\right] ,  \label{physical masses}
\end{eqnarray}%
where $\varepsilon $ is the sign function.

Written now directly in terms of the physical fields, the cubic interaction
terms are expressed as follows:%
\begin{equation}
U_{\mathrm{cub}}=\frac{\lambda _{0}^{\left( 3\right) }}{2}S_{0}^{2}h+\frac{%
\eta _{01}^{\left( 3\right) }}{2}S_{0}^{2}S_{1}+\frac{\lambda ^{\left(
3\right) }}{6}h^{3}+\frac{\eta _{1}^{\left( 3\right) }}{6}S_{1}^{3}+\frac{%
\lambda _{1}^{\left( 3\right) }}{2}h^{2}S_{1}+\frac{\lambda _{2}^{\left(
3\right) }}{2}hS_{1}^{2},  \label{U_cub}
\end{equation}%
where the cubic physical coupling constants are related to the original
parameters via the following relations:%
\begin{eqnarray}
\lambda _{0}^{\left( 3\right) } &=&\lambda _{0}v\cos \theta +\eta
_{01}v_{1}\sin \theta ;  \notag \\
\eta _{01}^{\left( 3\right) } &=&\eta _{01}v_{1}\cos \theta -\lambda
_{0}v\sin \theta ;  \notag \\
\lambda ^{\left( 3\right) } &=&\lambda v\cos ^{3}\theta +\frac{3}{2}\lambda
_{1}\sin 2\theta \left( v_{1}\cos \theta +v\sin \theta \right) +\eta
_{1}v_{1}\sin ^{3}\theta ;  \notag \\
\eta _{1}^{\left( 3\right) } &=&\eta _{1}v_{1}\cos ^{3}\theta -\frac{3}{2}%
\lambda _{1}\sin 2\theta \left( v\cos \theta -v_{1}\sin \theta \right)
-\lambda v\sin ^{3}\theta ;  \label{physical cubic coeff} \\
\lambda _{1}^{\left( 3\right) } &=&\lambda _{1}v_{1}\cos ^{3}\theta +\frac{1%
}{2}\sin 2\theta \left[ \left( 2\lambda _{1}-\lambda \right) v\cos \theta
-\left( 2\lambda _{1}-\eta _{1}\right) v_{1}\sin \theta \right] -\lambda
_{1}v\sin ^{3}\theta ;  \notag \\
\lambda _{2}^{\left( 3\right) } &=&\lambda _{1}v\cos ^{3}\theta -\frac{1}{2}%
\sin 2\theta \left[ \left( 2\lambda _{1}-\eta _{1}\right) v_{1}\cos \theta
+\left( 2\lambda _{1}-\lambda \right) v\sin \theta \right] +\lambda
_{1}v_{1}\sin ^{3}\theta .  \notag
\end{eqnarray}%
Also,  in terms of the physical fields, the quartic interactions  are
given by:%
\begin{eqnarray}
U_{\mathrm{quar}} &=&\frac{\eta _{0}}{24}S_{0}^{4}+\frac{\lambda ^{\left(
4\right) }}{24}h^{4}+\frac{\eta _{1}^{\left( 4\right) }}{24}S_{1}^{4}+\frac{%
\lambda _{0}^{\left( 4\right) }}{4}S_{0}^{2}h^{2}+\frac{\eta _{01}^{\left(
4\right) }}{4}S_{0}^{2}S_{1}^{2}+\frac{\lambda _{01}^{\left( 4\right) }}{2}%
S_{0}^{2}hS_{1}  \notag \\
&&+\frac{\lambda _{1}^{\left( 4\right) }}{6}h^{3}S_{1}+\frac{\lambda
_{2}^{\left( 4\right) }}{4}h^{2}S_{1}^{2}+\frac{\lambda _{3}^{\left(
4\right) }}{6}hS_{1}^{3},  \label{U_quar}
\end{eqnarray}%
where the physical quartic coupling constants are written in terms of the
original parameters of the theory as follows:%
\begin{eqnarray}
\lambda ^{\left( 4\right) } &=&\lambda \cos ^{4}\theta +\frac{3}{2}\lambda
_{1}\sin ^{2}2\theta +\eta _{1}\sin ^{4}\theta ;  \notag \\
\eta _{1}^{\left( 4\right) } &=&\eta _{1}\cos ^{4}\theta +\frac{3}{2}\lambda
_{1}\sin ^{2}2\theta +\lambda \sin ^{4}\theta ;  \notag \\
\lambda _{0}^{\left( 4\right) } &=&\lambda _{0}\cos ^{2}\theta +\eta
_{01}\sin ^{2}\theta ;  \notag \\
\eta _{01}^{\left( 4\right) } &=&\eta _{01}\cos ^{2}\theta +\lambda _{0}\sin
^{2}\theta ;  \notag \\
\lambda _{01}^{\left( 4\right) } &=&\frac{1}{2}\left( \eta _{01}-\lambda
_{0}\right) \sin 2\theta ;  \notag \\
\lambda _{1}^{\left( 4\right) } &=&\frac{1}{2}\left[ \left( 3\lambda
_{1}-\lambda \right) \cos ^{2}\theta -\left( 3\lambda _{1}-\eta _{1}\right)
\sin ^{2}\theta \right] \sin 2\theta ;  \notag \\
\lambda _{2}^{\left( 4\right) } &=&\lambda _{1}\cos ^{2}2\theta -\frac{1}{4}%
\left( 2\lambda _{1}-\eta _{1}-\lambda \right) \sin ^{2}2\theta ;  \notag \\
\lambda _{3}^{\left( 4\right) } &=&\frac{1}{2}\left[ \left( \eta
_{1}-3\lambda _{1}\right) \cos ^{2}\theta -\left( \lambda -3\lambda
_{1}\right) \sin ^{2}\theta \right] \sin 2\theta .
\label{physical quartic coeff}
\end{eqnarray}

Finally, after spontaneous breaking of the electroweak and $\mathbb{Z}_{2}$
symmetries, the part of the Standard Model lagrangian that is relevant to dark
matter annihilation writes, in terms of the physical fields $h$ and $S_{1}$,
as follows:%
\begin{eqnarray}
U_{\mathrm{SM}} &=&\sum_{f}\left( \lambda _{hf}h\bar{f}f+\lambda _{1f}S_{1}%
\bar{f}f\right) +\lambda _{hw}^{\left( 3\right) }hW_{\mu }^{-}W^{+\mu
}+\lambda _{1w}^{\left( 3\right) }S_{1}W_{\mu }^{-}W^{+\mu }  \notag \\
&&+\lambda _{hz}^{\left( 3\right) }h\left( Z_{\mu }\right) ^{2}+\lambda
_{1z}^{\left( 3\right) }S_{1}\left( Z_{\mu }\right) ^{2}+\lambda
_{hw}^{\left( 4\right) }h^{2}W_{\mu }^{-}W^{+\mu }+\lambda _{1w}^{\left(
4\right) }S_{1}^{2}W_{\mu }^{-}W^{+\mu }  \notag \\
&&+\lambda _{h1w}hS_{1}W_{\mu }^{-}W^{+\mu }+\lambda _{hz}^{\left( 4\right)
}h^{2}\left( Z_{\mu }\right) ^{2}+\lambda _{1z}^{\left( 4\right)
}S_{1}^{2}\left( Z_{\mu }\right) ^{2}+\lambda _{h1z}hS_{1}\left( Z_{\mu
}\right) ^{2}.  \label{SM interactions}
\end{eqnarray}%
The quantities $m_{f}$, $m_{w}$ and $m_{z}$ are the masses of the fermion $f$%
, the $W$ and the $Z$  gauge bosons respectively, and the above coupling constants are
given by the following relations:%
\begin{eqnarray}
\lambda _{hf} &=&-\frac{m_{f}}{v}\cos \theta ;\qquad \lambda _{1f}=\frac{%
m_{f}}{v}\sin \theta ;  \notag \\
\lambda _{hw}^{\left( 3\right) } &=&2\frac{m_{w}^{2}}{v}\cos \theta ;\qquad
\lambda _{1w}^{\left( 3\right) }=-2\frac{m_{w}^{2}}{v}\sin \theta ;  \notag
\\
\lambda _{hz}^{\left( 3\right) } &=&\frac{m_{z}^{2}}{v}\cos \theta ;\qquad
\lambda _{1z}^{\left( 3\right) }=-\frac{m_{z}^{2}}{v}\sin \theta ;  \notag \\
\lambda _{hw}^{\left( 4\right) } &=&\frac{m_{w}^{2}}{v^{2}}\cos ^{2}\theta
;\quad \lambda _{1w}^{\left( 4\right) }=\frac{m_{w}^{2}}{v^{2}}\sin
^{2}\theta ;\quad \lambda _{h1w}=-\frac{m_{w}^{2}}{v^{2}}\sin 2\theta ;
\notag \\
\lambda _{hz}^{\left( 4\right) } &=&\frac{m_{z}^{2}}{2v^{2}}\cos ^{2}\theta
;\quad \lambda _{1z}^{\left( 4\right) }=\frac{m_{z}^{2}}{2v^{2}}\sin
^{2}\theta ;\quad \lambda _{h1z}=-\frac{m_{z}^{2}}{2v^{2}}\sin 2\theta .
\label{SM couplings}
\end{eqnarray}

\section{Relic Density, Mutual Couplings and Perturbativity}

The original theory (\ref{U}) has nine parameters: three mass parameters $%
\left( \tilde{m}_{0},\mu ,\mu _{1}\right) $, three self-coupling constants $%
\left( \eta _{0},\lambda ,\eta _{1}\right) $ and three mutual coupling
constants $\left( \lambda _{0},\eta _{01},\lambda _{1}\right) $.
Perturbativity is assumed, hence all these original coupling constants are
small. The dark-matter self-coupling constant $\eta _{0}$ does not enter in the calculations of the
lowest-order processes of this work \cite{about-eta0}, so effectively, we
are left with eight parameters. The spontaneous breaking of the electroweak
and $\mathbb{Z}_{2}$ symmetries for the Higgs and $\chi _{1}$ fields
respectively introduces the two vacuum expectation values $v$ and $v_{1}$
given to lowest order in (\ref{vacua}). The value of $v$ is fixed
experimentally to be $246\mathrm{GeV}$  and for
the present work, we fix the value of $v_{1}$ at the order of the
electroweak scale, say $100\mathrm{GeV}$. Hence we are left with six
parameters. Four of these are chosen to be the three physical masses $m_{0}$
(dark matter), $m_{1}$ ($S_{1}$ field) and $m_{h}$ (Higgs), plus the mixing
angle $\theta $ between $S_{1}$ and $h$. We will fix the Higgs mass to $%
m_{h}=138\mathrm{GeV}$ and give, in this section, the mixing angle $\theta $
the two values $10^{\mathrm{o}}$ (small) and $40^{\mathrm{o}}$ (larger). The
two last parameters we choose are the two physical mutual coupling constants
$\lambda _{0}^{\left( 4\right) }$ (dark matter -- Higgs) and $\eta
_{01}^{\left( 4\right) }$ (dark matter -- $S_{1}$ particle), see (\ref%
{U_quar}).

In the framework of the thermal dynamics of the Universe within the standard cosmological model \cite{Kolb-Turner}, the WIMP relic density  is
related to its annihilation rate by the familiar relations:%
\begin{eqnarray}
\Omega _{D}\bar{h}^{2} &\simeq &\frac{1.07\times 10^{9}x_{f}}{\sqrt{g_{\ast }%
}m_{\mathrm{Pl}}\left\langle v_{12}\sigma _{\mathrm{ann}}\right\rangle
\mathrm{GeV}};  \notag \\
x_{f} &\simeq &\ln \frac{0.038m_{\mathrm{Pl}}m_{0}\left\langle v_{12}\sigma
_{\mathrm{ann}}\right\rangle }{\sqrt{g_{\ast }x_{f}}}.
\label{relic-density-relation}
\end{eqnarray}%
The notation is as follows: the quantity $\bar{h}$ is the Hubble constant in
units of 100km/(s$\times $Mpc), $m_{\mathrm{Pl}}=1.22\times 10^{19}\mathrm{%
GeV}$ the Planck mass, $m_{0}$ the dark matter mass, $x_{f}=m_{0}/T_{f}$ the
ratio of the dark matter mass to the freeze-out temperature $T_{f}$ and $%
g_{\ast }$ the number of relativistic degrees of freedom with a mass less
than $T_{f}$. The quantity $\left\langle v_{12}\sigma _{\mathrm{ann}%
}\right\rangle $ is the thermally averaged  annihilation cross-section of
a pair of two dark matter particles multiplied by their relative speed in
the center-of-mass reference frame.  Solving (\ref{relic-density-relation}) with the current value for the dark matter
relic density  $\Omega _{D}\bar{h}^{2}=0.105\pm 0.008$ \cite{WMAP2010} gives:
\begin{equation}
\left\langle v_{12}\sigma _{\mathrm{ann}}\right\rangle \simeq \left( 1.9\pm
0.2\right) \times 10^{-9}\mathrm{GeV}^{-2},  \label{v12sigma_annihi}
\end{equation}%
for a range of dark matter masses between roughly $10\mathrm{GeV}$ to $100%
\mathrm{GeV}$ and $x_{f}$ between 19.2 and 21.6, with about  0.4 thickness \cite%
{Weinberg}.

The value in (\ref{v12sigma_annihi}) for the dark matter annihilation
cross-section translates into a relation between the parameters of a given
theory entering the calculated expression of $\left\langle v_{12}\sigma _{%
\mathrm{ann}}\right\rangle $, hence imposing a constraint on these
parameters which will limit the intervals of possible dark matter masses.
This constraint can be exploited to examine aspects of the theory like
perturbativity. For example, in our model, we can obtain via (\ref%
{v12sigma_annihi}) the mutual coupling constant $\eta _{01}^{\left( 4\right)
}$ for given values of $\lambda _{0}^{\left( 4\right) }$, study its behavior
as a function of $m_{0}$ and tell which dark-matter mass regions are
consistent with perturbativity. Note that once the two mutual coupling
constants $\lambda _{0}^{\left( 4\right) }$ and $\eta _{01}^{\left( 4\right)
}$ are perturbative, all the other physical coupling constants will be. In
the study of this section, we choose the values $\lambda _{0}^{\left(
4\right) }=0.01$ (very weak), $0.2$ (weak) and 1 (large). We also let the
two masses $m_{0}$ and $m_{1}$ stretch from 0.1GeV to 120GeV, occasionally $%
m_{0}$ to 200GeV. Finally, note that we do not incorporate the uncertainty
in (\ref{v12sigma_annihi}) when imposing the relic-density constraint,
something that is sufficient in view of the descriptive nature of this work.
\begin{figure*}[h]
\begin{center}
\includegraphics[width=17.5cm,height=10.5cm]{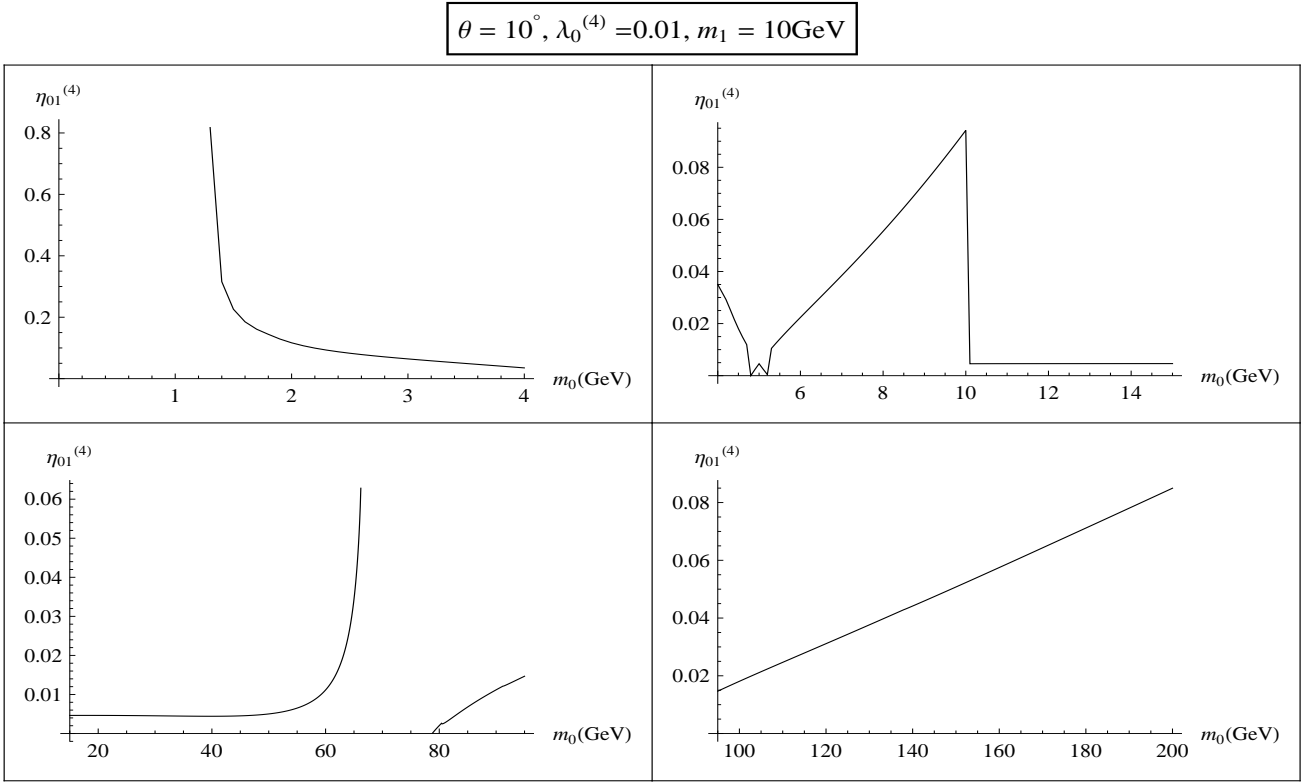}
\end{center}
\caption{\textit{{$\eta _{01}^{(4)}$
vs $m_{0}$ for small $m_1$, small mixing  and very small WIMP-Higgs coupling.}}}%
\label{eta014_theta-10_lambda04-001_m1-10}%
\end{figure*}
\begin{figure*}[h]
\begin{center}
\includegraphics[width=17.5cm,height=10cm]{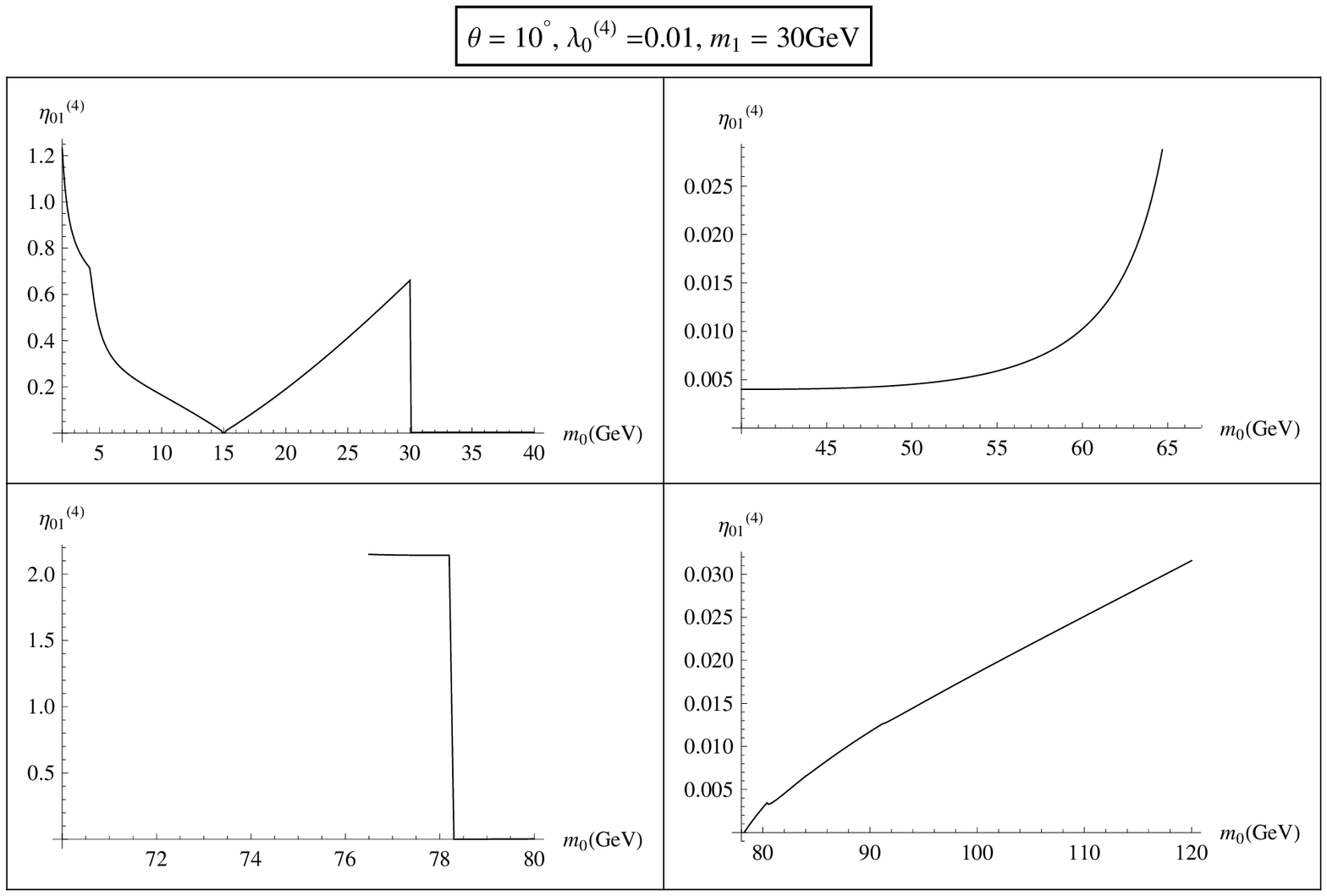}
\end{center}
\caption{\textit{{$\eta _{01}^{(4)}$
vs $m_{0}$ for moderate $m_{1}$, small mixing  and very small WIMP-Higgs coupling.}}}%
\label{eta014_theta-10_lambda04-001_m1-30}%
\end{figure*}
\begin{figure*}[h]
\begin{center}
\includegraphics[width=17.5cm,height=10cm]{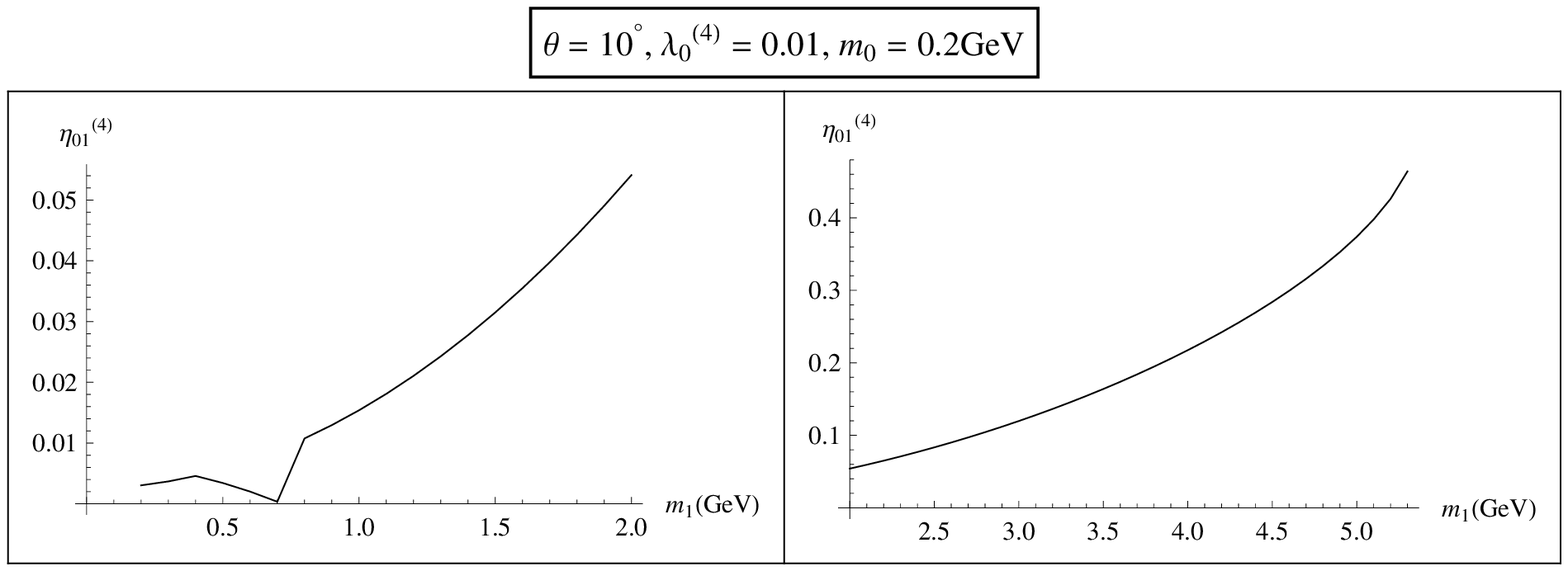}
\end{center}
\caption{\textit{{ $\eta _{01}^{(4)}$ vs $m_{1}$ for very light $S_1$, small mixing  and  very
small WIMP-Higgs coupling.}}}%
\label{eta014_theta-10_lambda04-001_m0-02}%
\end{figure*}

The dark matter annihilation cross sections (times the relative speed)
through all possible channels are given in the appendix. The quantity $%
\left\langle v_{12}\sigma _{\mathrm{ann}}\right\rangle $ is the sum of all
these contributions. Imposing $\left\langle v_{12}\sigma _{\mathrm{ann}%
}\right\rangle =1.9\times 10^{-9}\mathrm{GeV}^{-2}$ dictates the behavior of
$\eta _{01}^{\left( 4\right) }$, which is displayed as a function of the
dark matter mass $m_{0}$. Of course, as the parameters are numerous, the
behavior is bound to be rich and diverse. We cannot describe every bit of
it. Also, one has to note from the outset that for a given set of values for
the parameters, the solution to the relic-density constraint is not unique:
besides positive real solutions (when they exist), we may find negative real
or even complex solutions. It is beyond the scope of the present work to
investigate the nature and behavior of all the solutions. We are only
interested in finding the smallest positive real solution $\eta _{01}^{(4)}$
when it exists, looking at its behavior and finding out when it is small
enough to be perturbative.

\subsection{Small mixing angle and very weak dark matter -- Higgs coupling}
\begin{figure*}[h]
\begin{center}
\includegraphics[width=17.5cm,height=10cm]{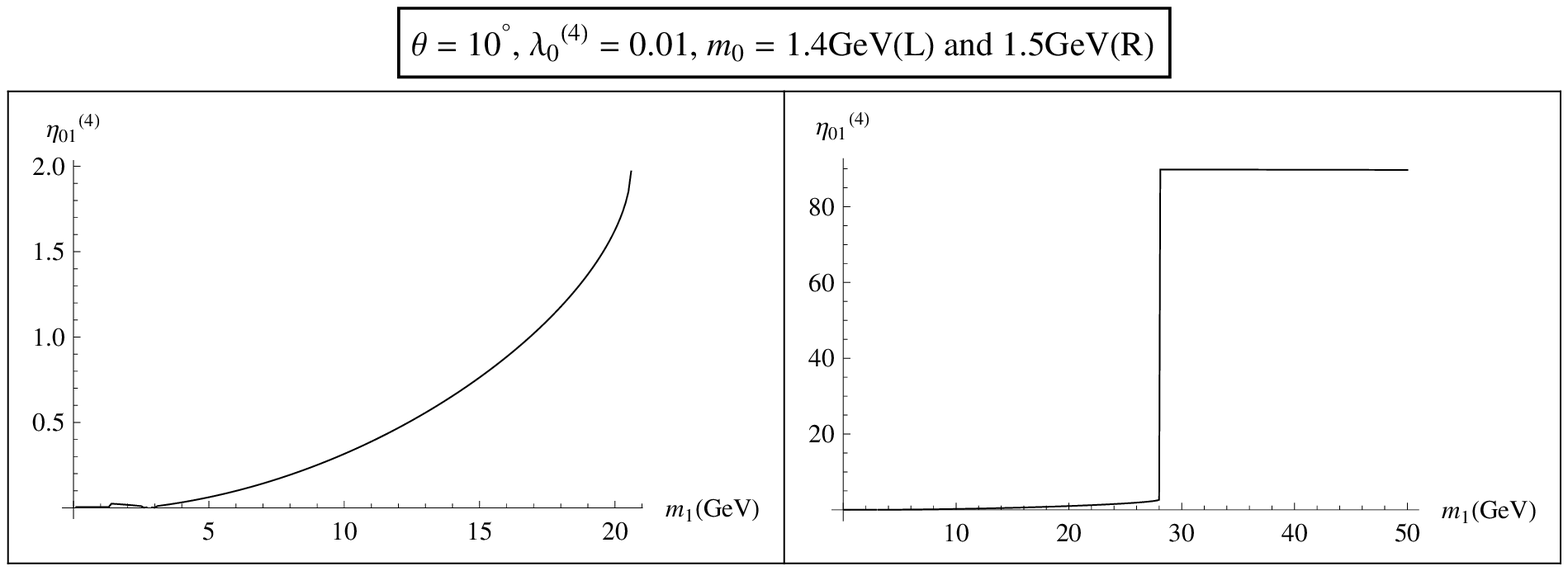}
\end{center}
\caption{\textit{{$\eta _{01}^{(4)}$
versus $m_{1}$ for $m_{0}$ above $\protect\tau $ threshold.}}}%
\label{eta014_theta-10_lambda04-001_m0-14-15}%
\end{figure*}
Let us describe briefly and only partly how the mutual $S_{0}$ -- $S_{1}$
coupling constant $\eta _{01}^{(4)}$ behaves as a function of the $S_{0}$
mass $m_{0}$. We start by a small mixing angle, say $\theta =10^{\mathrm{o}}$%
, and a very weak mutual $S_{0}$ -- Higgs coupling constant, say $\lambda
_{0}^{(4)}=0.01$. Let us also fix the $S_{1}$ mass first at the small value $%
m_{1}=10\mathrm{GeV}$. The corresponding behavior of $\eta _{01}^{(4)}$
versus $m_{0}$ is shown in Fig. \ref{eta014_theta-10_lambda04-001_m1-10}.%
\begin{figure*}[h]
\begin{center}
\includegraphics[width=17.5cm,height=10cm]{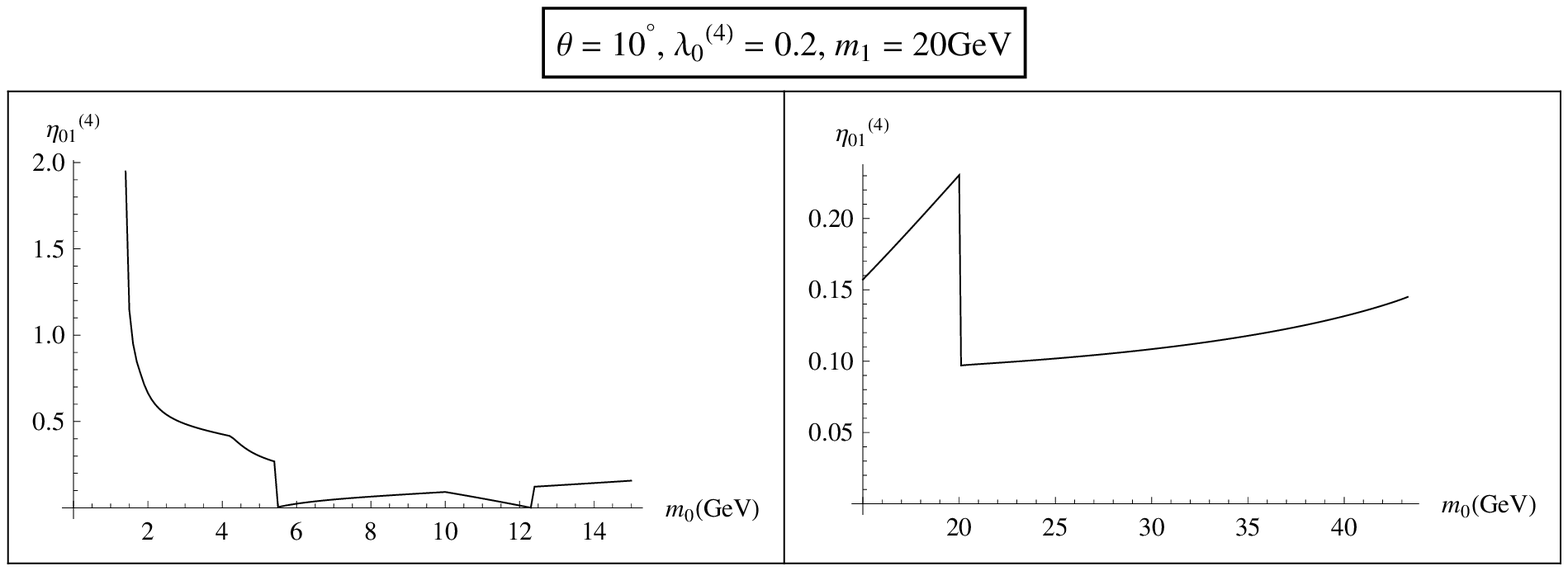}
\end{center}
\caption{\textit{{$\eta _{01}^{(4)}$ vs $m_{0}$ for small mixing, moderate $m_1$ and WIMP-Higgs coupling.}}}%
\label{eta014_theta-10_lambda04-02_m1-20}%
\end{figure*}

The range of $m_{0}$ shown is from $0.1\mathrm{GeV}$ to $200\mathrm{GeV}$, cut
in four intervals to allow for `local' features to be displayed\footnote{%
A logplot in this descriptive study is not advisable.}. We see that the
relic-density constraint on $S_{0}$ annihilation has no positive real
solution for $m_{0}\lesssim 1$.$3\mathrm{GeV}$, and so, with these very
small masses, $S_{0}$ cannot be a dark matter candidate. In other words, for
$m_{1}=10\mathrm{GeV}$, the particle $S_{0}$ cannot annihilate into the
lightest fermions only; inclusion of the $c$-quark is necessary. Note that
right about $m_{0}\simeq 1.3\mathrm{GeV}$, the $c$ threshold, the mutual
coupling constant $\eta _{01}^{(4)}$ starts at about $0.8$, a value, while
perturbative, that is roughly eighty-two-fold larger than the mutual $S_{0}$
-- Higgs coupling constant $\lambda _{0}^{(4)}$. Then $\eta _{01}^{(4)}$
decreases, steeply first, more slowly as we cross the $\tau $ mass towards
the $b$ mass. Just before $m_{1}/2$, the coupling $\eta _{01}^{(4)}$ hops
onto another solution branch that is just emerging from negative territory,
gets back to the first one at precisely $m_{1}/2$ as this latter carries now
smaller values, and then jumps up again onto the second branch as the first
crosses the $m_{0}$-axis down. It goes up this branch with a moderate slope
until $m_{0}$ becomes equal to $m_{1}$, a value at which the $S_{1}$
annihilation channel opens. Right beyond $m_{1}$, there is a sudden fall to
a value $\eta _{01}^{(4)}\simeq 0.0046$ that is about half the value of $%
\lambda _{0}^{(4)}$, and $\eta _{01}^{(4)}$ stays flat till $m_{0}\simeq 45%
\mathrm{GeV}$ where it starts increasing, sharply after 60GeV. In the mass interval
$m_{0}\simeq 66 \mathrm{GeV}-  79 \mathrm{GeV}$, there is a desert with no positive real solutions to the
relic-density constraint, hence no viable dark matter candidate.
Beyond $m_{0}\simeq 79 \mathrm{GeV}$, the mutual coupling constant $\eta _{01}^{(4)}$
keeps increasing monotonously, with a small notch at the $W$ mass and a less
noticeable one at the $Z$ mass. Note that for this value of $m_{1}\mathrm{\ }
$($10\mathrm{GeV}$), all values reached by $\eta _{01}^{(4)}$ in the mass
range considered, however large or small with respect to $\lambda _{0}^{(4)}$%
, are perturbativily acceptable.

Increasing $m_{1}$ to moderate values does not change the above qualitative
features. As an illustration, Fig. \ref{eta014_theta-10_lambda04-001_m1-30}
shows the behavior of $\eta _{01}^{(4)}$ as a function of $m_{0}$ for $%
m_{1}=30\mathrm{GeV}$, keeping the mixing angle $\theta =10^{\mathrm{o}}$,
still small, and the mutual $S_{0}$ -- Higgs coupling constant $\lambda
_{0}^{(4)}=0.01$, still very weak. The first thing to note is that not all values of $\eta _{01}^{(4)}$ are
perturbative. Indeed, $\eta _{01}^{(4)}$ does not start until $m_{0}\simeq
1.5\mathrm{GeV}$, but with the very large value\footnote{%
This feature is not displayed in figure \ref%
{eta014_theta-10_lambda04-001_m1-30} to avoid masking the other much smaller
values taken by the mutual coupling.} 89.8. It decreases very sharply right
after, to 2.04 at about 1.6GeV. It continues to decrease with a pronounced\
change in the slope at the $b$ threshold. Effects at the masses$\ m_{1}/2$
and $m_{1}$ similar to those of figure \ref%
{eta014_theta-10_lambda04-001_m1-10} do occur here too. There is a desert
 that lies in this case in the mass interval 66.5GeV -- 76.5GeV. At the
upper bound, the coupling $\eta _{01}^{(4)}$ takes the value 2.15 and
decreases very slowly till $m_{0}\simeq 78.2\mathrm{GeV}$. Right after this
mass, it plunges down to catch up with a solution branch that is just
emerging from negative values. This solution branch increases steadily with
two small notches at the $W$ and $Z$ masses. A similar global behavior
occurs at other moderate $m_{1}$ masses, with varying local features.

Because of the very-small-$m_{0}$ deserts described and visible on Fig. %
 \ref{eta014_theta-10_lambda04-001_m1-10}, one may ask whether the model ever
allows for very light dark matter. To look into this, we fix $m_{0}$ at small values and let $m_{1}$ vary. Take first $m_{0}=0.2\mathrm{GeV}$ and
see Fig. \ref{eta014_theta-10_lambda04-001_m0-02}. The
allowed $S_{0}$ annihilation channels are the very light fermions $e,u,d,\mu
$ and $s$, plus $S_{1}$ when $m_{1}<m_{0}$. Note that we still have $\theta
=10^{\mathrm{o}}$ and $\lambda _{0}^{(4)}=0.01$. Qualitatively, we notice
that in fact, there are no solutions for $m_{1}<0.2\mathrm{GeV}$ ($=m_{0}$
here), a mass at which $\eta _{01}^{\left( 4\right) }$ takes the very small
value $\simeq 0.003$. It goes up a solution branch and leaves it at $%
m_{1}\simeq 0.4\mathrm{GeV}$ to descend on a second branch that enters
negative territory at $m_{1}\simeq 0.7 \mathrm{GeV}$, forcing $\eta
_{01}^{(4)}$ to return onto the first branch. There is an accelerated
increase till $m_{1}\simeq 5 \mathrm{GeV}$, a value at which $\eta
_{01}^{(4)}\simeq 0.5$. And then a desert, no positive real solutions, no
viable dark matter.

Increasing $m_{0}$ until about $1.3 \mathrm{GeV}$ does not change these
overall features: some `movement' for very small values of $m_{1}$ and then
an accelerated increase till reaching a desert with a lower bound that
changes with $m_{0}$. For example, the desert starts at $m_{1}\simeq 6.8%
\mathrm{GeV}$ for $m_{0}=0.6 \mathrm{GeV}$ and $m_{1}\simeq 7.3 \mathrm{GeV}$
for $m_{0}=1.2 \mathrm{GeV}$. Note that in all these cases where $%
m_{0}\lesssim 1.3 \mathrm{GeV}$, all values of $\eta _{01}^{(4)}$ are
perturbative. Therefore, the model can very well accommodate very light dark
matter with a restricted range of $S_{1}$ masses.

However, the situation changes after the inclusion of the $\tau $
annihilation channel. Indeed, as Fig. \ref%
{eta014_theta-10_lambda04-001_m0-14-15} shows, for $m_{0}=1.4 \mathrm{GeV}$,
though the overall shape of the behavior of $\eta _{01}^{(4)}$ as a function
of $m_{1}$ is qualitatively the same, the desert threshold is pushed
significantly higher, to $m_{1}\simeq 20 \mathrm{GeV}$. But more
significantly, $\eta _{01}^{(4)}$ starts to be larger than one already at $%
m_{1}\simeq 17 \mathrm{GeV}$, therefore loosing perturbativity. For $%
m_{0}=1.5\mathrm{GeV}$, the desert is effectively erased as we have a sudden
jump to highly non-perturbative values of $\eta _{01}^{(4)}$ right after $%
m_{1}\simeq 28 \mathrm{GeV}$. Such a behavior stays with larger values of $%
m_{0}$. But for $m_{1}\lesssim 20 \mathrm{GeV}$ (case $m_{0}=1.5 \mathrm{GeV}$%
), the values of $\eta _{01}^{(4)}$ are smaller than one and physical use of
the model is possible if needed.
\begin{figure*}[h]
\begin{center}
\includegraphics[width=17.5cm,height=10cm]{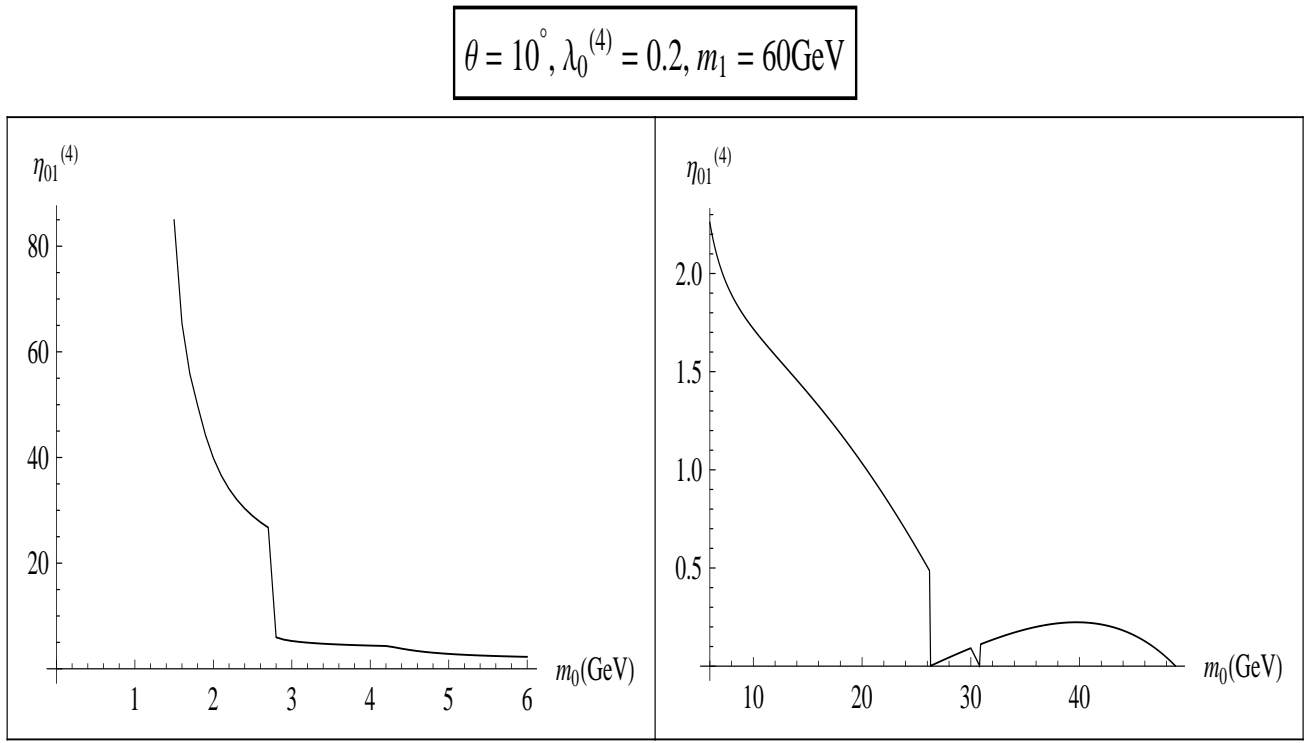}
\end{center}
\caption{\textit{{$\eta _{01}^{(4)}$ versus $m_{0}$ for heavy $S_1$, small mixing  and small WIMP-Higgs coupling.}}}%
\label{eta014_theta-10_lambda04-02_m1-60}%
\end{figure*}

\subsection{Small mixing angle and larger dark matter -- Higgs couplings}

What are the effects of the relic-density constraint when we vary the
parameter $\lambda _{0}^{(4)}$? Let us keep the Higgs -- $S_{1}$ mixing
angle small ($\theta =10^{\mathrm{o}}$) and increase $\lambda _{0}^{(4)}$,
first to $0.2$ and later to 1. For $\lambda _{0}^{(4)}=0.2$, Figure \ref%
{eta014_theta-10_lambda04-02_m1-20} shows the behavior of $\eta _{01}^{(4)}$
as a function of the dark matter mass $m_{0}$ when $m_{1}=20 \mathrm{GeV}$.
We see that $\eta _{01}^{(4)}$ starts at $m_{0}\simeq 1.4 \mathrm{GeV}$ with
a value of about $1.95$. It decreases with a sharp change of slope at the $b$
threshold, then makes a sudden dive at about 5 GeV, a change of branch at $%
m_{1}/2$ down till about $12 \mathrm{GeV}$ where it jumps up back onto the
previous branch just before going to cross into negative territory. It drops
sharply at $m_{0}=m_{1}$ and then increases slowly until $m_{0}\simeq 43.3%
\mathrm{GeV}$. Beyond, there is nothing, a desert.

This is of course different from the situation of very small $\lambda _{0}^{(4)}$
like in Fig. \ref{eta014_theta-10_lambda04-001_m1-10} and Fig. \ref%
{eta014_theta-10_lambda04-001_m1-30} above: here we see some kind of natural
dark-matter mass `confinement' to small-moderate viable\footnote{%
Note that the values of $\eta _{01}^{(4)}$ for $1.6 \mathrm{GeV}\lesssim
m_{0}\lesssim 43.3 \mathrm{GeV}$ are all perturbative.} values.

Still for $\lambda _{0}^{(4)}=0.2$ with $m_{1}=60 \mathrm{GeV}$ this time,
Fig. \ref{eta014_theta-10_lambda04-02_m1-60} shows $\eta _{01}^{(4)}$
starting very high ($\simeq 85 \mathrm{GeV})$ at $m_{0}\simeq 1.5\mathrm{GeV}$, decreasing
quickly  with a first sudden drop at 2.7 GeV and a second one to zero at
26.2 GeV. A solution branch is then picked up -- left briefly at $m_{1}/2$ --
until 49 GeV and then nothing. What is peculiar here is that, in contrast
with previous situations, the desert starts at a mass $m_{0}<m_{1}$, i.e.,
before the opening of the $S_{1}$ annihilation channel. In other words, the
dark matter is annihilating into the light fermions only and the model is
perturbatively viable in the range 20GeV -- 49GeV.
\begin{figure*}[h]
\begin{center}
\includegraphics[width=16.5cm,height=9cm]{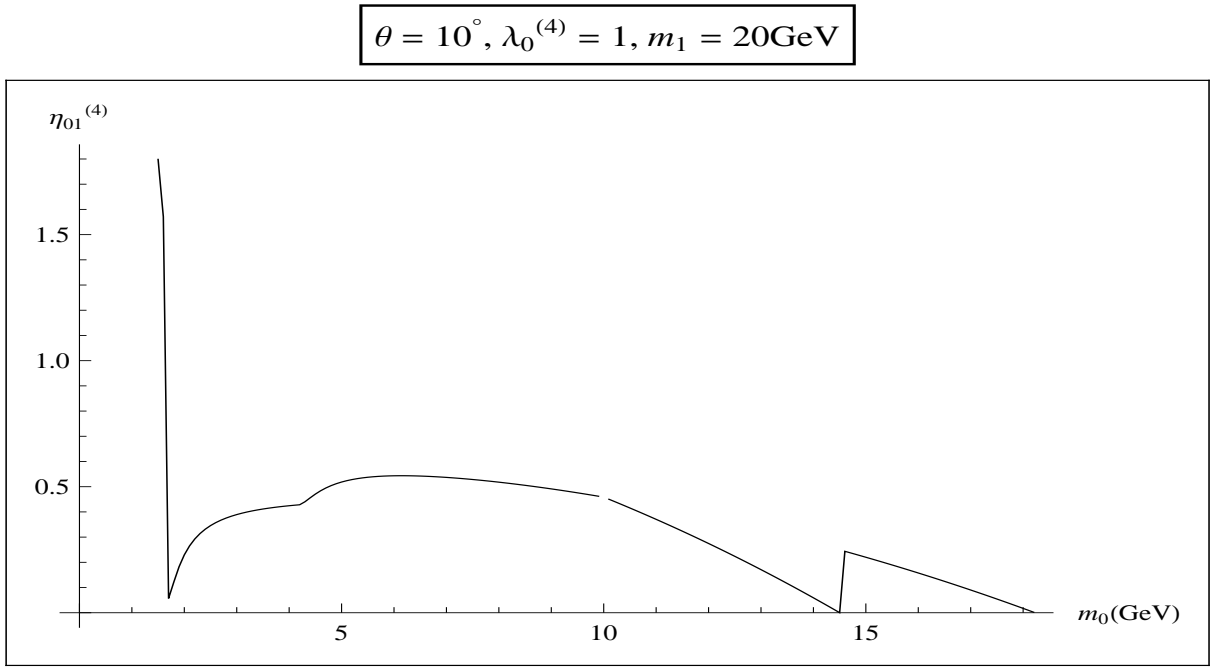}
\end{center}
\caption{\textit{{$\eta %
_{01}^{(4)}$ versus $m_{0}$ for medium  $m_1$, small mixing and large WIMP-higgs coupling.}}}%
\label{eta014_theta-10_lambda04-1_m1-20}%
\end{figure*}

The case $\lambda _{0}^{(4)}=1$ with $m_{1}=20\mathrm{GeV}$ is displayed in
Fig. \ref{eta014_theta-10_lambda04-1_m1-20}. There are no solutions below $%
m_{0}\simeq 1.5\mathrm{GeV}$ at which $\eta _{01}^{(4)}\simeq 1.80$. From
this value, $\eta _{01}^{(4)}$ slips down very quickly to pick up less
abruptly when crossing the $\tau $ threshold. There is a significant change
in the slope at the crossing of the $b$ mass. Note the absence of a solution
at $m_{1}/2$, which is a new feature, present for other values of $m_{1}$
not displayed here. Beyond $m_{1}/2$, there is a slight change in the
downward slope, a change of solution branch, and that goes until $14.5%
\mathrm{GeV}$ where $\eta _{01}^{(4)}$ jumps to catch up with the previous
branch. It goes down this branch until about $18 \mathrm{GeV}$ where the
desert starts.

\begin{figure*}[h]
\begin{center}
\includegraphics[width=16.5cm,height=9cm]{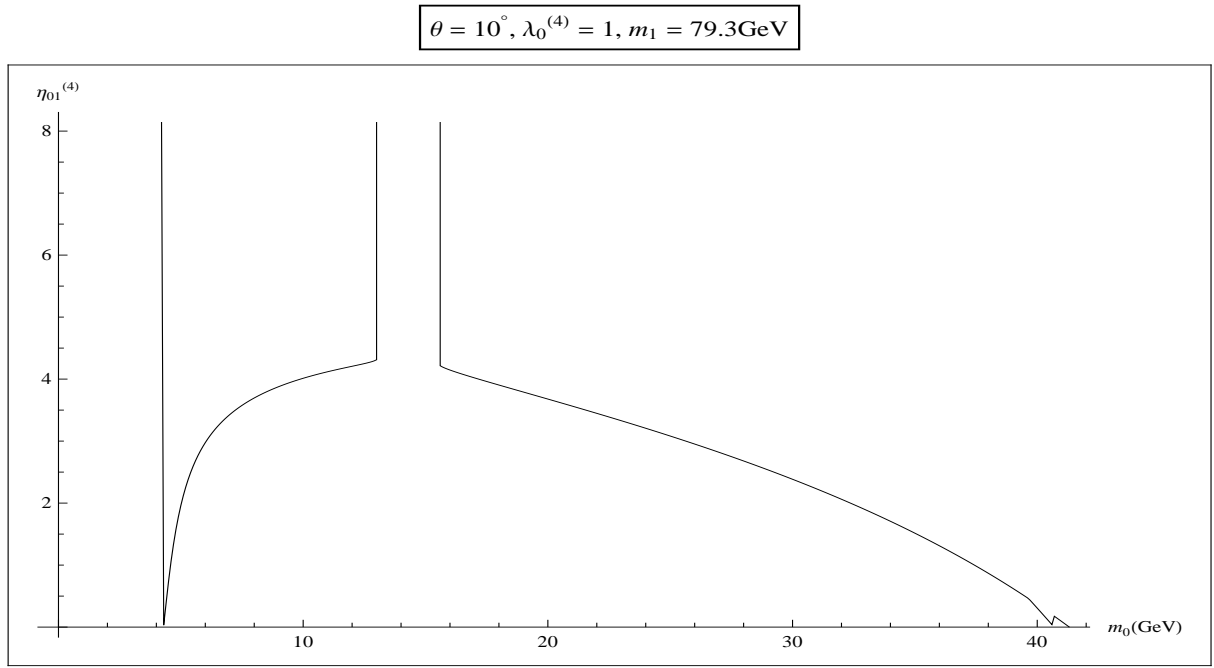}
\end{center}
\caption{\textit{{$\eta %
_{01}^{(4)}$ versus $m_{0}$ for heavy $S_1$, small mixing and large WIMP-Higgs coupling.}}}%
\label{eta014_theta-10_lambda04-1_m1-793}%
\end{figure*}
We have studied the behavior of $\eta _{01}^{(4)}$ as a function of $m_{0}$
for other values of $m_{1}$ between $20\mathrm{GeV}$ and $100\mathrm{GeV}$
while keeping $\theta =10^{\mathrm{o}}$ and $\lambda _{0}^{(4)}=1$. For $%
m_{1}\lesssim 79.2\mathrm{GeV}$, the behavior is qualitatively quite similar
to that shown in Fig. \ref{eta014_theta-10_lambda04-1_m1-20}, but beyond
this mass, there is a highly non-perturbative branch $\eta _{01}^{(4)}$
jumps  onto at small and moderate values of $m_{0}$. This highly
non-perturbative region stretches in size as $m_{1}$ increases. Fig. \ref%
{eta014_theta-10_lambda04-1_m1-793} displays this new feature. Note that on
this figure, not all of the range of $\eta _{01}^{(4)}$ is shown in order to
allow the small-coupling regions to be displayed; the high values of $\eta
_{01}^{(4)}$ are in the two thousands. Note also that it is the same highly
non-perturbative solution branch $\eta _{01}^{(4)}$\ jumps onto for other
large values of $m_{1}$.

\subsection{Larger mixing angles}

Last in this descriptive study is to see the effects of larger values of the
$S_{1}$ -- Higgs mixing angle $\theta $. We give it here the value $\theta
=40^{\mathrm{o}}$ and tune back the mutual $S_{0}$ -- Higgs coupling
constant $\lambda _{0}^{(4)}$ to the very small value $0.01$. Figure \ref%
{eta014_theta-40_lambda04-001_m1-20} shows the behavior of $\eta _{01}^{(4)}$
as a function of $m_{0}$ for $m_{1}=20\mathrm{GeV}$. One recognizes features
similar to those of the case $\theta =10^{\mathrm{o}}$, though coming in
different relative sizes. The very-small-$m_{0}$ desert ends at about $0.3%
 \mathrm{GeV}$. There are by-now familiar features at the $c$ and $b$ masses,
$m_{1}/2$ and $m_{1}$. Two relatively small forbidden intervals (deserts)
appear for relatively large values of the dark matter mass: $67.3 \mathrm{GeV}%
\ -70.9 \mathrm{GeV}$ and $79.4 \mathrm{GeV}-90.8\mathrm{GeV}$. The $W$ mass region
is forbidden but there is action as we cross the $Z$ mass.
\begin{figure*}[h]
\begin{center}
\includegraphics[width=17.5cm,height=10cm]{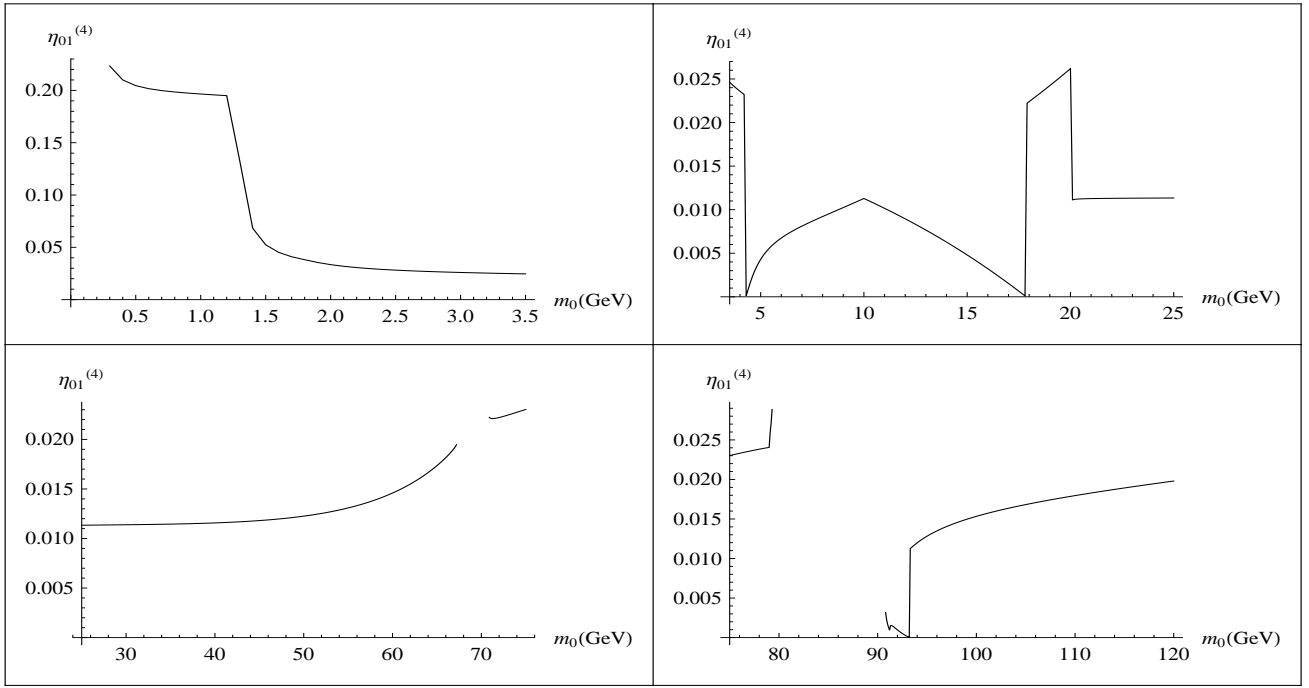}
\end{center}
\caption{\textit{{$\eta _{01}^{(4)}$
versus $m_{0}$ for moderate $m_1$, moderate mixing and small WIMP-Higgs coupling.}}}%
\label{eta014_theta-40_lambda04-001_m1-20}%
\end{figure*}

\begin{figure*}[h]
\begin{center}
\includegraphics[width=17.5cm,height=10cm]{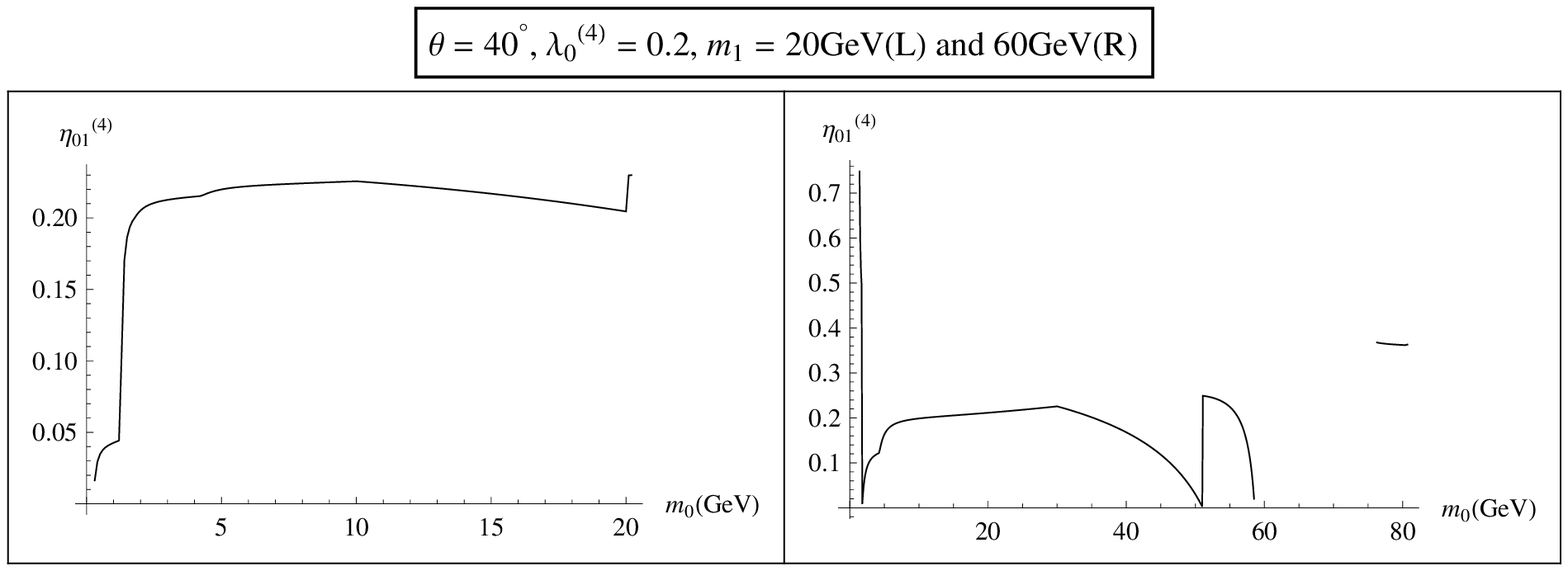}
\end{center}
\caption{\textit{{$\eta _{01}^{(4)}$ versus $m_{0}$ for moderate (L) and large (R) $m_1$, large mixing and moderate WIMP-Higgs coupling.}}}%
\label{eta014_theta-40_lambda04-02_m1-20-60}%
\end{figure*}
\begin{figure*}[h]
\begin{center}
\includegraphics[width=17.5cm,height=8cm]{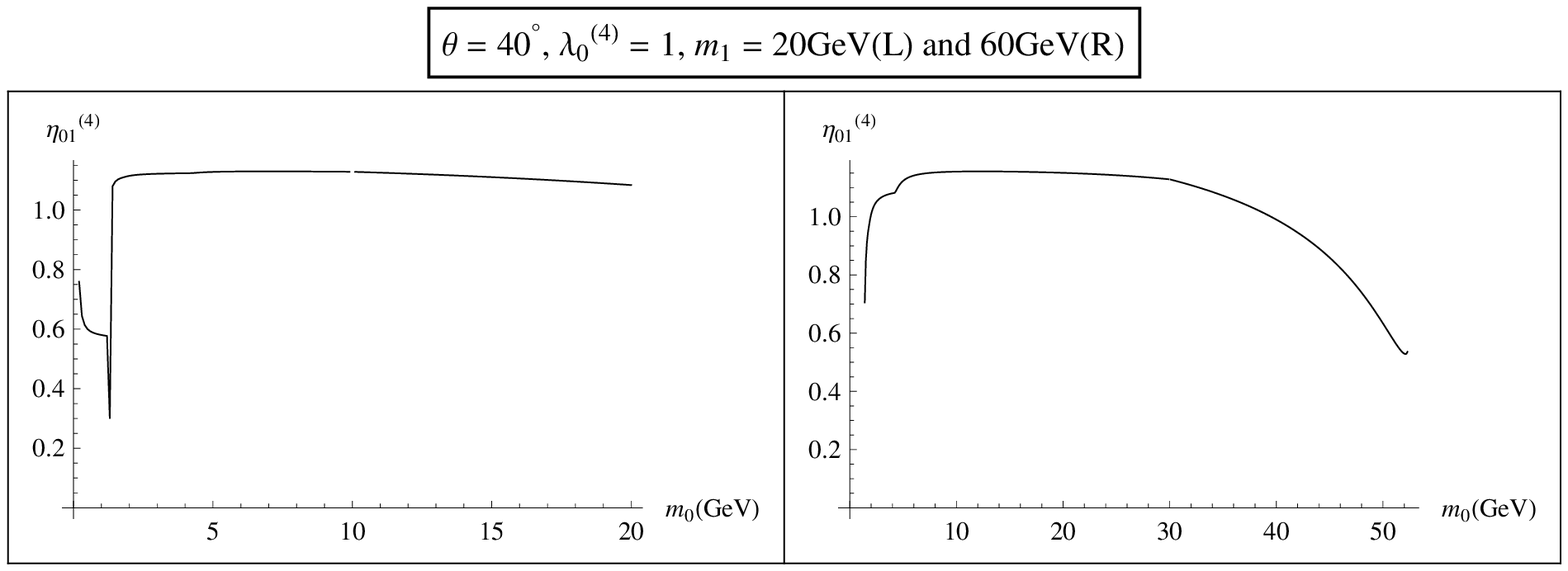}
\end{center}
\caption{\textit{{$\eta _{01}^{(4)}$ versus $m_{0}$ for moderate (L) and large (R) $m_1$, large  mixing and large WIMP-Higgs coupling.}}}%
\label{eta014_theta-40_lambda04-1_m1-20-60}%
\end{figure*}
 Other values of $%
m_{1}$, not displayed because of space, behave similarly with an additional
effect, namely,  a sudden drop in slope at $m_{0}=(m_{h}+m_{1})/2$ coming from the
ignition of $S_{0}$ annihilation into $S_{1}$ and Higgs.
We have also worked out the cases $\lambda _{0}^{(4)}=0.2$ and 1 for $\theta
=40^{\mathrm{o}}$. The case $\lambda _{0}^{(4)}=0.2$ is displayed in Fig. %
\ref{eta014_theta-40_lambda04-02_m1-20-60} and presents differences with the
corresponding small-mixing situation $\theta =10^{\mathrm{o}}$. Indeed, for $%
m_{1}=20\mathrm{GeV}$, the first feature we notice is a smoother behavior;
compare with Fig. \ref{eta014_theta-10_lambda04-02_m1-20}. Here, $\eta
_{01}^{(4)}$ starts at $m_{0}\simeq 0.3\mathrm{GeV}$ with the small value $%
\simeq 0.016$ and goes up, faster at the $c$ mass and with a small effect at
the $b$ mass. It increases very slowly until $m_{1}/2$ and decreases very
slowly until $m_{0}=m_{1}$, and then there is a sudden change of branch
followed immediately by a desert\footnote{%
Except for the very tiny interval $78.5\mathrm{GeV}-79.0\mathrm{GeV}$ not
displayed on Fig. \ref{eta014_theta-40_lambda04-02_m1-20-60}.}. So here
too the model naturally confines the mass of a viable dark matter to
small-moderate values, a dark matter particle annihilating into light
fermions only. What is also noticeable is that there is stability of $\eta
_{01}^{(4)}$ around the value of $\lambda _{0}^{(4)}$ in the interval $1.5%
\mathrm{GeV}-20\mathrm{GeV}$ $(=m_{1}$ here$)$.

The case $m_{1}=60\mathrm{GeV}$ presents also overall similarities as well
as noticeable differences with the corresponding case $\theta =10^{\mathrm{o}%
}$, see Fig. \ref{eta014_theta-10_lambda04-02_m1-60}. The first difference
is that all values of $\eta _{01}^{(4)}$ are perturbative. This latter
starts at $m_{0}\simeq 1.4\mathrm{GeV}$ with the value $\sim 0.75$, goes
down and jumps to catch up with another solution branch emerging from
negative territory when crossing the $\tau $ mass. It increases, kicking up
when crossing the $b$-quark mass. It changes slope down at $m_{1}/2$ and
goes to zero at about $51\mathrm{GeV}$. It jumps up onto another branch that
goes down to zero also at about $58.6\mathrm{GeV}$, just below $m_{1}$, and
then there is a desert, except for the small interval $76.3\mathrm{GeV}-80.5%
\mathrm{GeV}$.

The case $\lambda _{0}^{(4)}=1$ is shown in Fig. \ref%
{eta014_theta-40_lambda04-1_m1-20-60}. Global similarities with the previous
case are apparent. All values of $\eta _{01}^{(4)}$ are perturbative and the
mass range is naturally confined to the interval $0.2\mathrm{GeV}-20\mathrm{%
GeV}$ for $m_{1}=20\mathrm{GeV}$, and $1.4\mathrm{GeV-}52.3\mathrm{GeV}$ for
$m_{1}=60\mathrm{GeV}$. We note action at the usual masses and, in
particular, we see there are no solutions at $m_{0}=m_{1}/2$ like in the
case $\theta =10^{\mathrm{o}}$. We note here too the quasi-constancy of $%
\eta _{01}^{(4)}$ for most of the available range.

Finally, we note that we have worked out larger mixing angles, notably $%
\theta =75^{\mathrm{o}}$. In general, these cases do not display any new
features worth discussing: the overall behavior is mostly similar to what we
have seen, with expected relative variations in size.

\section{Dark-Matter Direct Detection}

Experiments like CDMS II \cite{cdmsII-1,cdmsII-2}, XENON 10/100 \cite%
{xenon10-1,xenon100}, DAMA/LIBRA \cite{dama-libra} and CoGeNT \cite{cogent}
search directly for a dark matter signal. Such a signal would typically come
from the elastic scattering of a dark matter WIMP off a non-relativistic
nucleon target. However, throughout the years, such experiments have not yet
detected an unambiguous signal, but rather yielded increasingly stringent
exclusion bounds on the dark matter -- nucleon
elastic-scattering total cross section $\sigma _{\det }$ in terms of the
dark matter mass $m_{0}$.

In order for a theoretical dark-matter model to be viable, it has to satisfy these bounds. It is therefore natural to inquire
whether the model we present in this work has any capacity of describing
dark matter. Hence, we have to calculate $\sigma _{\det }$ as a function of $%
m_{0}$ for different values of the parameters ($\theta ,\lambda
_{0}^{(4)},m_{1})$ and project its behavior against the experimental bounds.
We will limit ourselves to the region 0.1GeV -- 100GeV as we are interested
in light dark matter. As experimental bounds, we will use the results from
CDMSII and XENON100, as well as the future projections of SuperCDMS \cite%
{supercdms} and XENON1T \cite{xenon1t}. The results of CoGeNT, DAMA/LIBRA and CRESST
will be discussed elsewhere \cite{abada-nasri-1}. As the figures below show
\cite{exp-data}, in the region of our interest, XENON100 is only slightly
tighter than CDMSII, SuperCDMS significantly lower and XENON1T the most
stringent by far. But it is important to note that all these results loose
reasonable predictability in the very light sector, say below 5GeV.

The scattering of $S_0$ off a SM fermion $f$ occurs via the t-channel exchange of the SM higgs  and $S_1$. In the non-relativistic limit, the effective Lagrangian describing this interaction reads
\begin{eqnarray}
{\cal L}^{(eff)}_{S_0-f} = a_f \bar{f} f S_0^2,
\end{eqnarray}
where
\begin{eqnarray}
a_f = -\frac{m_f}{2v}\left[ \frac{\lambda _{0}^{\left( 3\right) }\cos{\theta}}{m_{h}^{2}}-\frac{\eta _{01}^{\left( 3\right) }\sin{\theta}}{m_{1}^{2}%
}\right].
\end{eqnarray}
In this case the total cross section for this process is given by :%
\begin{equation}
\sigma _{S_{0}f\rightarrow S_{0}f}=\frac{m_{f}^{4}}{4\pi \left(
m_{f}+m_{0}\right) ^{2}v^2}\left[ \frac{\lambda _{0}^{\left( 3\right) }\cos{\theta}}{m_{h}^{2}}-\frac{\eta _{01}^{\left( 3\right) }\sin{\theta}}{m_{1}^{2}%
}\right] ^{2}.  \label{s_0 f cross-section}
\end{equation}%
\newline
At the nucleon level, the effective interaction between a nucleon $N = p$ or $n$  and $S_0$ has the form
\begin{eqnarray}
{\cal L}^{(eff)}_{S_0-N} = a_N \bar{N} N S_0^2,
\end{eqnarray}
where the effective nucleon-$S_0$ coupling constants is given by:%
\begin{equation}
a_N = \frac{\left(m_{N}-\frac{7}{9}m_{B}\right)}{v}\left[ \frac{\lambda _{0}^{\left( 3\right) }\cos{\theta}}{m_{h}^{2}}-\frac{\eta _{01}^{\left( 3\right) }\sin{\theta}}{m_{1}^{2}%
}\right].
\label{nucleon h and S_0 coupling}
\end{equation}%
In this relation, $m_{N}$ is the nucleon mass and $m_{B}$ the baryon mass in
the chiral limit \cite{he-li-li-tandean-tsai}. The total cross section for
non-relativistic $S_{0}$ -- $N$ elastic scattering is therefore:%
\begin{equation}
\sigma _{\det }\equiv \sigma _{S_{0}N\rightarrow S_{0}N}=\frac{%
m_{N}^{2}\left( m_{N}-\frac{7}{9}m_{B}\right) ^{2}}{4\pi \left(
m_{N}+m_{0}\right) ^{2}v^2}\left[ \frac{\lambda _{0}^{\left( 3\right) }\cos{\theta}}{%
m_{h}^{2}} - \frac{\eta _{01}^{\left( 3\right) }\sin{\theta}}{m_{1}^{2}}\right] ^{2}.
\label{s_0 N cross-section}
\end{equation}
\indent The rest of this section is devoted to a brief discussion of the behavior of
$\sigma _{\det }$ as a function of $m_{0}$. We will of course impose the
relic-density constraint on the dark matter annihilation cross section (\ref%
{v12sigma_annihi}). But in addition, we will require that the coupling
constants are perturbative, and so impose the additional requirement $0\leq
\eta _{01}^{(4)}\leq 1$. Also, here too, the choices of the sets of values
of the parameters ($\theta ,\lambda _{0}^{(4)},m_{1}$) can by no means be
exhaustive but only indicative. Furthermore, though a detailed description
of the behavior of $\sigma _{\det }$ could be interesting in its own right,
we will refrain from doing so in this work as there is no need for it, and
content ourselves with mentioning overall features and trends. Generally,
as $m_{0}$ increases, the detection cross section $\sigma _{\det }$ starts
from high values, slopes down to minima that depend on the parameters and
then picks up moderately. There are features and action at the usual mass
thresholds, with varying sizes and shapes. Excluded regions are there, those
coming from the relic-density constraint and new ones originating from the
additional perturbativity requirement. Close to the upper boundary of the
mass interval considered in this study, there is no universal behavior to
mention as in some cases $\sigma _{\det }$ will increase monotonously and,
in some others, it will decrease or `not be there' at all. Let us finally
remark that the logplots below may not show these general features clearly
as these latter are generally distorted.
\begin{figure*}[h]
\begin{center}
\includegraphics[width=17.5cm,height=8cm]{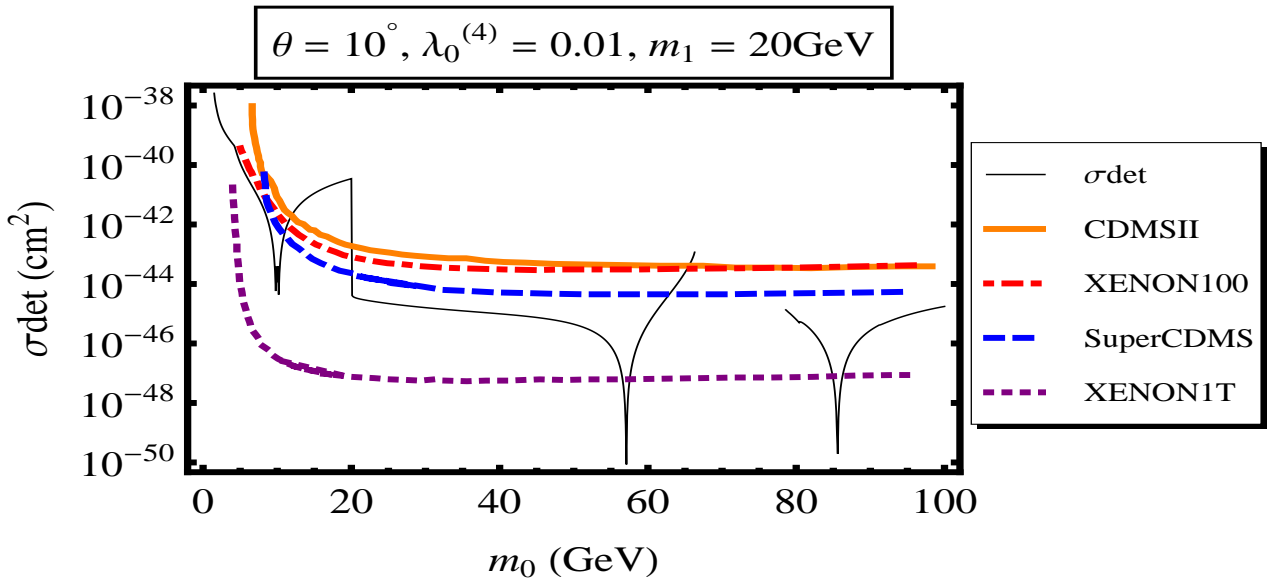}
\end{center}
\caption{\textit{{Elastic $N-S_0$ scattering cross-section  as a function of $m_0$ for moderate $m_1$, small mixing and small WIMP-Higgs coupling.}}}%
\label{exp_det_theta-10_lambda04-001_m1-20}%
\end{figure*}
\begin{figure*}[h]
\begin{center}
\includegraphics[width=17.5cm,height=8cm]{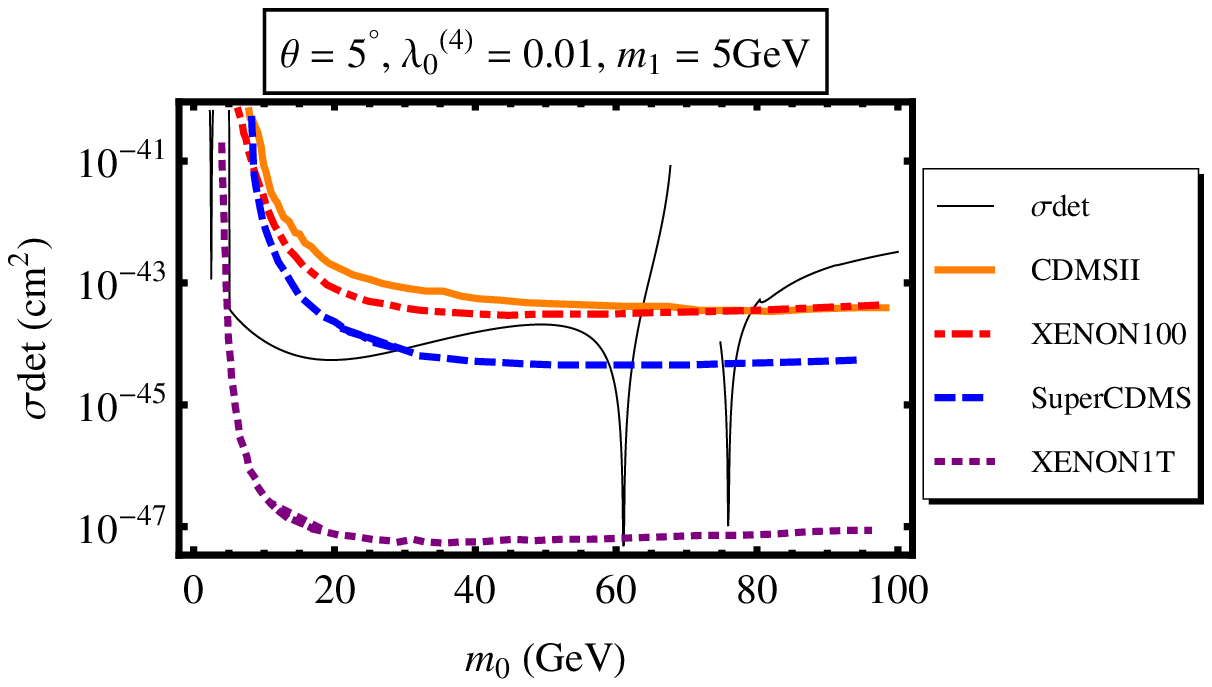}
\end{center}
\caption{\textit{{Elastic $N-S_0$ scattering cross-section  as a function of $m_0$  for light $S_1$, small mixing and small WIMP-Higgs coupling.}}}%
\label{exp_det_theta-05_lambda04-001_m1-05}%
\end{figure*}
\begin{figure*}[h]
\begin{center}
\includegraphics[width=17.5cm,height=8cm]{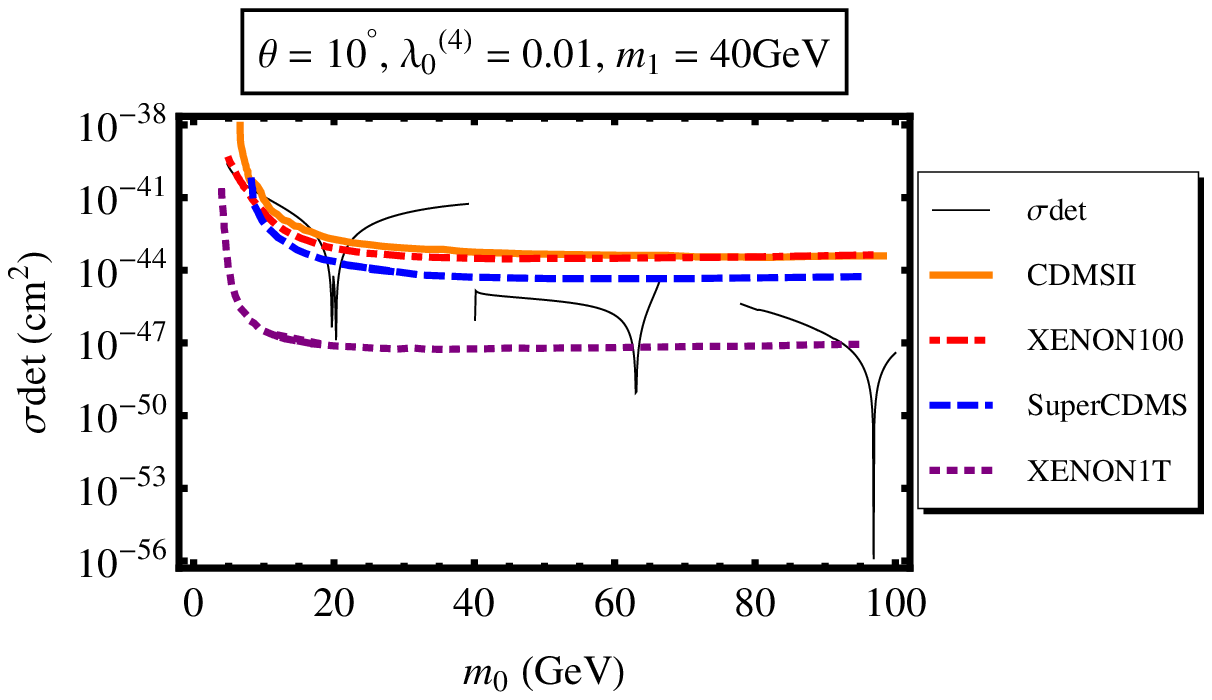}
\end{center}
\caption{\textit{{Elastic $N-S_0$ scattering cross-section  as a function of $m_0$ for medium $m_1$, small mixing and small WIMP-Higgs coupling.}}}%
\label{exp_det_theta-10_lambda04-001_m1-40}%
\end{figure*}

Let us start with the small Higgs -- $S_{1}$ mixing angle $\theta =10^{%
\mathrm{o}}$ and the very weak mutual $S_{0}$ -- Higgs coupling $\lambda
_{0}^{(4)}=0.01$. Fig. \ref{exp_det_theta-10_lambda04-001_m1-20} shows the
behavior of $\sigma _{\det }$ versus $m_{0}$ in the case $m_{1}=20\mathrm{GeV%
}$. We see that for the two mass intervals  $20\mathrm{GeV}-65\mathrm{GeV}$ and  $75\mathrm{GeV}-100\mathrm{GeV}$,  plus an almost singled-out peaks at  $m_{0}=m_{1}/2$,  the elastic scattering cross section is below  the projected  sensitivity of SuperCDMS. However,  XENON1T will probe all the these masses , except  $m_0 \simeq 58$ GeV and $85$ GeV.

Also, as we see in Fig. \ref{exp_det_theta-10_lambda04-001_m1-20}, most of the mass range for very
light dark matter is excluded for these values of the parameters. Is this
systematic? In general, smaller values of $m_{1}$ drive the predictability
ranges to the lighter sector of the dark matter masses. Figure \ref%
{exp_det_theta-05_lambda04-001_m1-05} illustrates this pattern. We have
taken  $m_{1}=5\mathrm{GeV}$, just above the lighter quarks threshold. In the small-mass region, we see that SuperCDMS is passed in the range $5%
\mathrm{GeV}-30\mathrm{GeV}$. Again , all this mass ranges will be probed by XENON1T experiment, except a sharp peak at $m_{0}=m_{1}/2=2.5\mathrm{GeV}$, but for such a very light mass, the experimental results are not without ambiguity.

\begin{figure*}[h]
\begin{center}
\includegraphics[width=17.5cm,height=8cm]{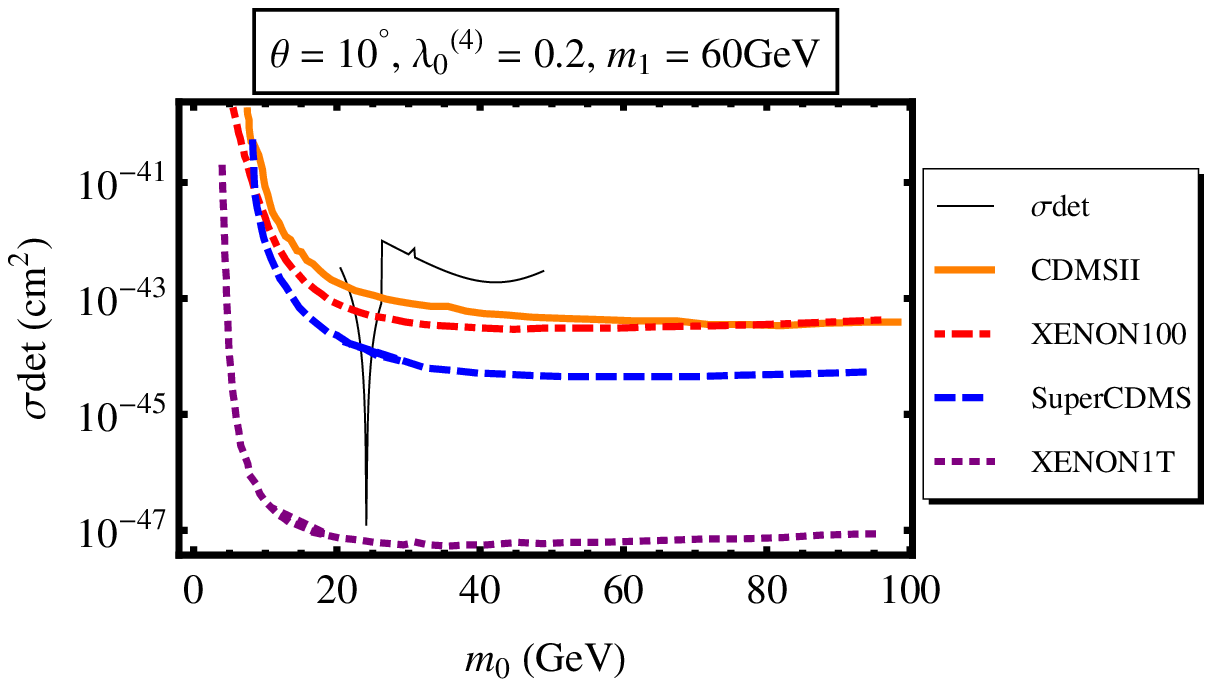}
\end{center}
\caption{\textit{{Elastic $N-S_0$ scattering cross-section  as a function of $m_0$ for heavy $S_1$, small mixing and moderate WIMP-Higgs coupling.}}}%
\label{exp_det_theta-10_lambda04-02_m1-60}%
\end{figure*}
\begin{figure*}[h]
\begin{center}
\includegraphics[width=17.5cm,height=8cm]{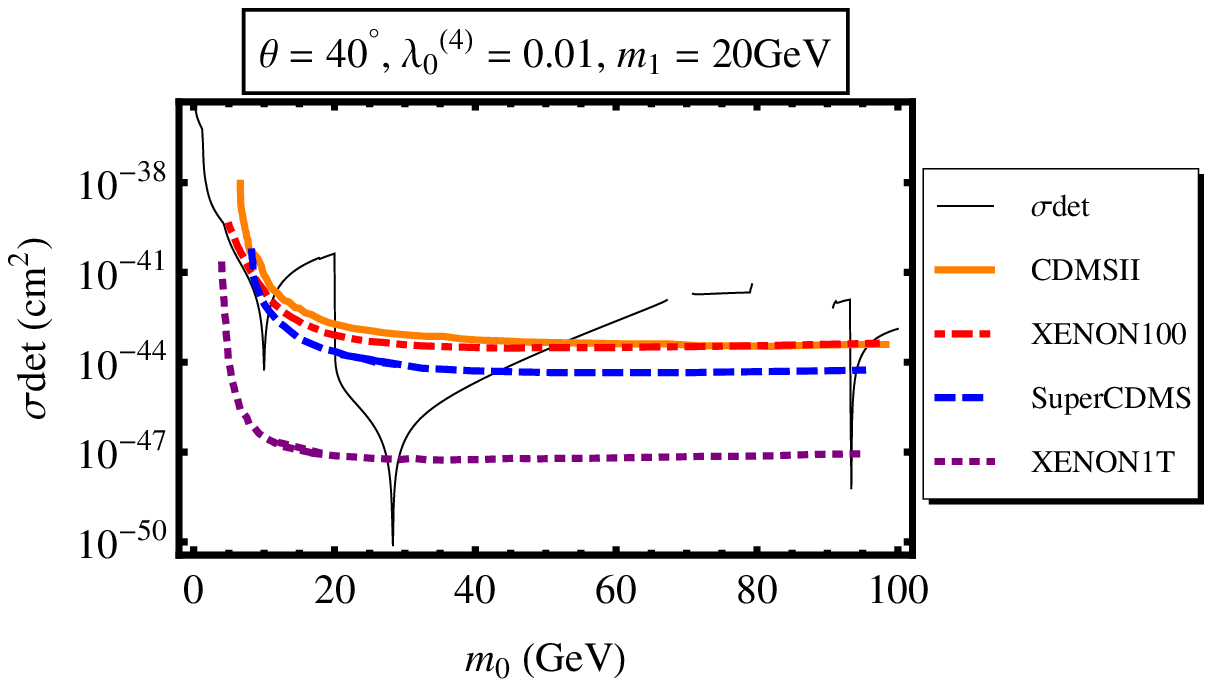}
\end{center}
\caption{\textit{{Elastic $N-S_0$ scattering cross-section  as a function of $m_0$ for moderate $m_1$, large mixing and small WIMP-Higgs coupling.}}}%
\label{exp_det_theta-40_lambda04-001_m1-20}%
\end{figure*}
\begin{figure*}[h]
\begin{center}
\includegraphics[width=17.5cm,height=8cm]{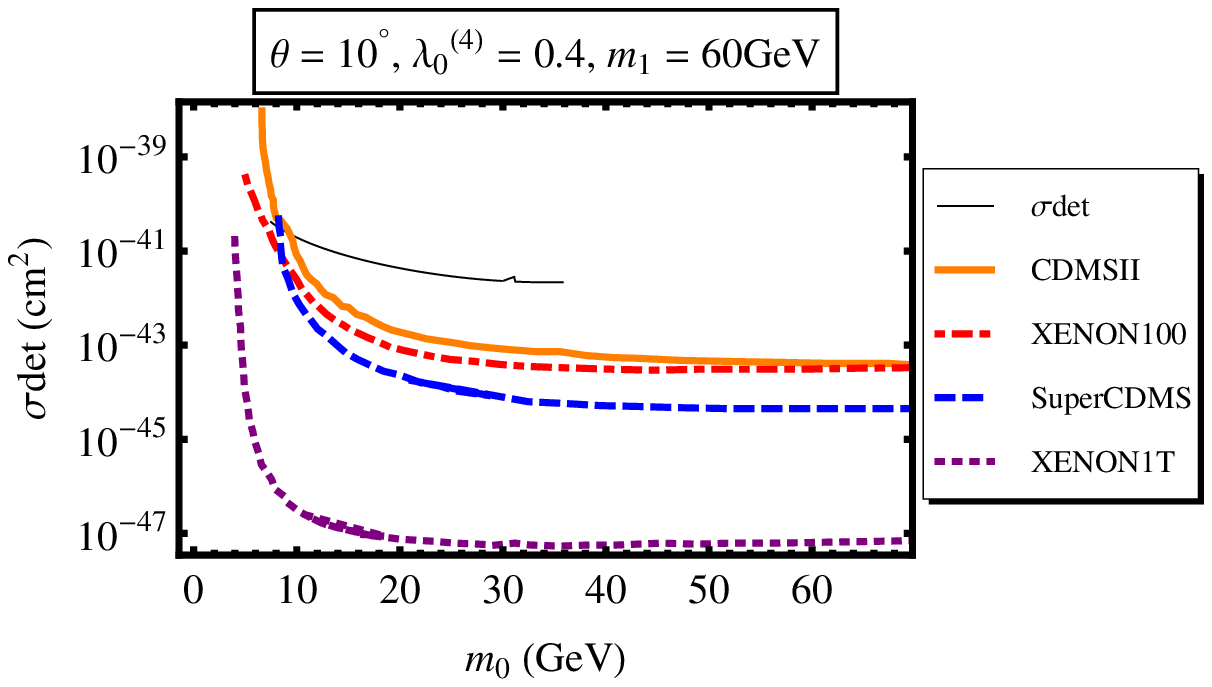}
\end{center}
\caption{\textit{{elastic cross-section $\sigma_{el}$  as a function of $S_0$ mass for heavy $S_1$, small mixing and relatively large WIMP-Higgs coupling.}}}%
\label{exp_det_theta-10_lambda04-04_m1-60}%
\end{figure*}
Reversely, increasing $m_{1}$ shuts down possibilities for very light dark
matter and thins the intervals as it drives the predicted masses to larger values. For instance,  in  figure \ref%
{exp_det_theta-10_lambda04-001_m1-40} where $m_{1}=40\mathrm{GeV}$, in
addition to the peak at $m_{1}/2$ that crosses SuperCDMS but not XENON1T, we
see acceptable masses in the ranges 40GeV -- 65GeV and 78GeV -- up. Here too
the intervals narrow as we descend, surviving XENON1T as spiked peaks at
62 GeV and around 95 GeV.

A larger  mutual coupling constant $\lambda _{0}^{(4)}$ has the general
effect of squeezing the acceptable intervals of $m_{0}$ by pushing the
values of $\sigma _{\det }$ up. As an illustration, see figure \ref%
{exp_det_theta-10_lambda04-02_m1-60} where we have taken $\lambda
_{0}^{(4)}=0.2$ and a larger value of $m_{1}=60\mathrm{GeV}$. In this
example, already XENON100 excludes all the masses below 100 GeV except a relatively narrow peak  at $m_{1}/2$.

Increasing the mixing angle $\theta $ has also the general effect of
increasing the value of $\sigma _{\det }$. Figure \ref%
{exp_det_theta-40_lambda04-001_m1-20} shows this trend for $\theta =40^{%
\mathrm{o}}$; compare with Fig. \ref{exp_det_theta-10_lambda04-001_m1-20}.
The only allowed masses by the current bounds of CDMSII and XENON100  are  between $20$ GeV and $50$ GeV, the  narrow interval around $%
m_{1}/2$, and another very sharp one, at about 94GeV.   The projected sensitivity of XENON1T will  probe all  mass ranges except those at  $m_{0}\simeq 30\mathrm{GeV}$ and 94GeV.

Finally, it happens that there are regions of the parameters for which the
model has no predictability. See figure \ref%
{exp_det_theta-10_lambda04-04_m1-60} for illustration. We have combined the
effects of increasing the values of the two parameters $\lambda^{(4)}_0$ and $m_1$. As we see, we
barely get something at $m_{1}/2$ that cannot even cross XENON100 down to
SuperCDMS.

\section{Concluding remarks}

In this work, we  presented a plausible scenario for light cold dark matter, (for masses lighter than $100$ GeV). This latter consists in enlarging the Standard Model with two gauge-singlet $\mathbb{Z}_2$-symmetric scalar fields. One is the dark matter field $S_0$, stable, while the other undergoes spontaneous symmetry breaking, resulting in the physical field $S_1$. This opens additional channels through which $S_0$ can annihilate, hence a reducing its number density. The model is parametrized by three quantities: the physical mutual coupling constant $\lambda _{0}^{(4)}$
between $S_{0}$  and the Higgs, the mixing angle $\theta $ between $S_{1}$ and the Higgs  and the mass $m_{1}$ of the particle $S_{1}$. We first imposed on $S_{0}$ annihilation cross section  the constraint from the observed dark-matter relic density and studied its effects through the behavior of the physical mutual coupling constant $\eta _{01}^{(4)}$ between $S_{0}$ and
$S_{1}$ as a function of the dark matter mass $m_{0}$. Apart from forbidden
regions (deserts) or others where perturbativity is lost, we find that for
most values of the three parameters, there are viable solutions in the
small-moderate masses of the dark matter sector. Deserts are found for most of the ranges of the
parameters whereas perturbativity is lost mainly for larger values of $m_{1}$%
. Through the behavior of $\eta _{01}^{(4)}$, we could see the mass
thresholds which mostly affect the annihilation of dark matter, and these are at
the $c$, $\tau $ and $b$ masses, as well as $m_{1}/2$ and $m_{1}$.

The current experimental bounds from CDMSII and XENON100 put a strong constraint on the $S_0$ masses in the range between $10$ to $20$ GeV. For small values of $m_1$, very light dark matter is viable, with a mass as small as one  GeV. This is  of course useful for understanding the results of  the experiments DAMA/LIBRA, CoGeNT ,  CRESST \cite{CRESST} as well as the recent data of the Fermi Gamma Ray Space Telescope \cite{HG} .  The projected sensitivity of future WIMP direct searches such as XENON1T will probe all the $S_0$ masses between $5$ GeV and $100$ GeV.

The next step to take is to test the model against the  phenomenological constraints.
Indeed, one important feature of the model is that it mixes the $S_{1}$ field with the
Higgs. This must have implications on the Higgs detection through the
measurable channels. Current experimental bounds from LEP II data can be used to constrain
our mixing angle $\theta $, and possibly other parameters. In addition, a  very light $S_0$ and/or  $S_1$  will contribute to the invisible decay  of $J/\psi$ and $\Upsilon$ mesons and can lead to a significant branching fraction. These constraints
can be injected back into the model and restrain further its domain of
validity. These issues are under current investigation \cite{abada-nasri-1}.

Also, in this work, the $S_{1}$ vacuum expectation value $v_{1}$ was taken equal
to 100GeV, but a priori, nothing prevents us from
considering other scales. However, taking $v_1$ much larger than the electro-weak scale  requires  $\eta^{(4)}_{01}$ to be very tiny , which will result in the suppression of  the crucial  annihilation channel $S_0S_0 \rightarrow S_1S_1$ . Also, we have
fixed the Higgs mass to $m_{h}=138\mathrm{GeV}$, which is consistent with
the current acceptable experimental bounds \cite{particle-data-group}. Yet,
it can be useful to ask here too what the effect of changing this mass would
be.

Finally, in this study, besides the dark matter field $S_{0}$, only one
extra field has been considered. Naturally, one can generalize the
investigation to include $N$ such fields and discuss the cosmology and
particle phenomenology in terms of $N$. It just happens that the model is
rich enough to open new possibilities in the quest of dark matter worth
pursuing.

\appendix

\section{Dark matter annihilation cross-sections}

The cross-sections related to the annihilation $S_{0}$ into the scalar
particles are as follows. For the $hh$ channel, we have:
\begin{eqnarray}
v_{12}\sigma _{S_{0}S_{0}\rightarrow hh} &=&\frac{\sqrt{m_{0}^{2}-m_{h}^{2}}%
}{64\pi m_{0}^{3}}\Theta (m_{0}-m_{h})\left[ \left( \lambda
_{0}^{(4)}\right) ^{2}+\frac{4\lambda _{0}^{(4)}\left( \lambda
_{0}^{(3)}\right) ^{2}}{m_{h}^{2}-2m_{0}^{2}}+\frac{2\lambda
_{0}^{(4)}\lambda _{0}^{(3)}\lambda ^{(3)}}{4m_{0}^{2}-m_{h}^{2}}\right.
\notag \\
&&+\frac{2\lambda _{0}^{(4)}\lambda _{1}^{(3)}\eta
_{01}^{(3)}(4m_{0}^{2}-m_{1}^{2})}{(4m_{0}^{2}-m_{1}^{2})^{2}+\epsilon
_{1}^{2}}+\frac{4\left( \lambda _{0}^{(3)}\right) ^{4}}{%
(m_{h}^{2}-2m_{0}^{2})^{2}}+\frac{4\lambda ^{(3)}\left( \lambda
_{0}^{(3)}\right) ^{3}}{(4m_{0}^{2}-m_{h}^{2})(m_{h}^{2}-2m_{0}^{2})}  \notag
\\
&&+\frac{4\left( \lambda _{0}^{(3)}\right) ^{2}\lambda _{1}^{(3)}\eta
_{01}^{(3)}(4m_{0}^{2}-m_{1}^{2})}{\left[ (4m_{0}^{2}-m_{1}^{2})^{2}+%
\epsilon _{1}^{2}\right] (m_{h}^{2}-2m_{0}^{2})}+\frac{\left( \lambda
^{(3)}\right) ^{2}\left( \lambda _{0}^{(3)}\right) ^{2}}{%
(4m_{0}^{2}-m_{h}^{2})^{2}}  \notag \\
&&+\left. \frac{\left( \lambda _{1}^{(3)}\right) ^{2}\left( \eta
_{01}^{(3)}\right) ^{2}}{(4m_{0}^{2}-m_{1}^{2})^{2}+\epsilon _{1}^{2}}+\frac{%
2\lambda _{0}^{(3)}\lambda _{1}^{(3)}\lambda ^{(3)}\eta
_{01}^{(3)}(4m_{0}^{2}-m_{1}^{2})}{\left[ (4m_{0}^{2}-m_{1}^{2})^{2}+%
\epsilon _{1}^{2}\right] (4m_{0}^{2}-m_{h}^{2})}\right] .
\end{eqnarray}%
The $\Theta $ function is the step function. For the $S_{1}S_{1}$ channel,
we have the result:%
\begin{eqnarray}
v_{12}\sigma _{S_{0}S_{0}\rightarrow S_{1}S_{1}} &=&\frac{\sqrt{%
m_{0}^{2}-m_{1}^{2}}}{64\pi m_{0}^{3}}\Theta (m_{0}-m_{1})\left[ \left( \eta
_{01}^{(4)}\right) ^{2}+\frac{4\eta _{01}^{(4)}\left( \eta
_{01}^{(3)}\right) ^{2}}{m_{1}^{2}-2m_{0}^{2}}+\frac{2\eta _{01}^{(4)}\eta
_{01}^{(3)}\eta _{1}^{(3)}}{4m_{0}^{2}-m_{1}^{2}}\right.  \notag \\
&&+\frac{2\eta _{01}^{(4)}\lambda _{0}^{(3)}\lambda
_{2}^{(3)}(4m_{0}^{2}-m_{h}^{2})}{(4m_{0}^{2}-m_{h}^{2})^{2}+\epsilon
_{h}^{2}}+\frac{4\left( \eta _{01}^{(3)}\right) ^{4}}{%
(m_{1}^{2}-2m_{0}^{2})^{2}}+\frac{4\left( \eta _{01}^{(3)}\right) ^{3}\eta
_{1}^{(3)}}{(4m_{0}^{2}-m_{1}^{2})(m_{1}^{2}-2m_{0}^{2})}  \notag \\
&&+\frac{4\left( \eta _{01}^{(3)}\right) ^{2}\lambda _{0}^{(3)}\lambda
_{2}^{(3)}(4m_{0}^{2}-m_{h}^{2})}{\left[ (4m_{0}^{2}-m_{h}^{2})^{2}+\epsilon
_{h}^{2}\right] (m_{1}^{2}-2m_{0}^{2})}+\frac{\left( \eta _{01}^{(3)}\right)
^{2}\left( \eta _{1}^{(3)}\right) ^{2}}{(4m_{0}^{2}-m_{1}^{2})^{2}}  \notag
\\
&&+\left. \frac{\left( \lambda _{0}^{(3)}\right) ^{2}\left( \lambda
_{2}^{(3)}\right) ^{2}}{(4m_{0}^{2}-m_{h}^{2})^{2}+\epsilon _{h}^{2}}+\frac{%
2\eta _{01}^{(3)}\eta _{1}^{(3)}\lambda _{0}^{(3)}\lambda
_{2}^{(3)}(4m_{0}^{2}-m_{h}^{2})}{\left[ (4m_{0}^{2}-m_{h}^{2})^{2}+\epsilon
_{h}^{2}\right] (4m_{0}^{2}-m_{1}^{2})}\right] .
\end{eqnarray}%
For the $hS_{1}$ channel, we have:%
\begin{eqnarray}
v_{12}\sigma _{S_{0}S_{0}\rightarrow S_{1}h} &=&\frac{\sqrt{\left[
4m_{0}^{2}-(m_{h}-m_{1})^{2}\right] \left[ 4m_{0}^{2}-(m_{h}+m_{1})^{2}%
\right] }}{128\pi m_{0}^{4}}\Theta (2m_{0}-m_{h}-m_{1})  \notag \\
&&\left[ \left( \lambda _{01}^{(4)}\right) ^{2}+\frac{8\lambda
_{01}^{(4)}\eta _{01}^{(3)}\lambda _{0}^{(3)}}{m_{h}^{2}+m_{1}^{2}-4m_{0}^{2}%
}+\frac{2\lambda _{01}^{(4)}\lambda _{0}^{(3)}\lambda _{1}^{(3)}}{%
4m_{0}^{2}-m_{h}^{2}}+\frac{2\lambda _{01}^{(4)}\eta _{01}^{(3)}\lambda
_{2}^{(3)}}{4m_{0}^{2}-m_{1}^{2}}\right.  \notag \\
&&+\frac{16\left( \eta _{01}^{(3)}\right) ^{2}\left( \lambda
_{0}^{(3)}\right) ^{2}}{\left( m_{h}^{2}+m_{1}^{2}-4m_{0}^{2}\right) ^{2}}+%
\frac{8\left( \lambda _{0}^{(3)}\right) ^{2}\eta _{01}^{(3)}\lambda
_{1}^{(3)}}{\left( m_{h}^{2}+m_{1}^{2}-4m_{0}^{2}\right)
(4m_{0}^{2}-m_{h}^{2})}  \notag \\
&&+\frac{8\left( \eta _{01}^{(3)}\right) ^{2}\lambda _{0}^{(3)}\lambda
_{2}^{(3)}}{\left( m_{h}^{2}+m_{1}^{2}-4m_{0}^{2}\right)
(4m_{0}^{2}-m_{1}^{2})}+\frac{\left( \lambda _{0}^{(3)}\right) ^{2}\left(
\lambda _{1}^{(3)}\right) ^{2}}{(4m_{0}^{2}-m_{h}^{2})^{2}}  \notag \\
&&+\left. \frac{2\eta _{01}^{(3)}\lambda _{0}^{(3)}\lambda _{1}^{(3)}\lambda
_{2}^{(3)}}{(4m_{0}^{2}-m_{h}^{2})(4m_{0}^{2}-m_{1}^{2})}+\frac{\left(
\lambda _{2}^{(3)}\right) ^{2}\left( \eta _{01}^{(3)}\right) ^{2}}{%
(4m_{0}^{2}-m_{1}^{2})^{2}}\right] .
\end{eqnarray}

The annihilation cross-section into fermions is:%
\begin{eqnarray}
v_{12}\sigma _{S_{0}S_{0}\rightarrow f\bar{f}} &=&\frac{\sqrt{\left(
m_{0}^{2}-m_{f}^{2}\right) ^{3}}}{4\pi m_{0}^{3}}\Theta \left(
m_{0}-m_{f}\right) \left[ \frac{\left( \lambda _{0}^{\left( 3\right)
}\lambda _{hf}\right) ^{2}}{\left( 4m_{0}^{2}-m_{h}^{2}\right) ^{2}+\epsilon
_{h}^{2}}+\frac{\left( \eta _{01}^{\left( 3\right) }\lambda _{1f}\right) ^{2}%
}{\left( 4m_{0}^{2}-m_{1}^{2}\right) ^{2}+\epsilon _{1}^{2}}\right.  \notag
\\
&&+\left. \frac{2\lambda _{0}^{\left( 3\right) }\eta _{01}^{\left( 3\right)
}\lambda _{hf}\lambda _{1f}\left( 4m_{0}^{2}-m_{h}^{2}\right) \left(
4m_{0}^{2}-m_{1}^{2}\right) }{\left[ \left( 4m_{0}^{2}-m_{h}^{2}\right)
^{2}+\epsilon _{h}^{2}\right] \left[ \left( 4m_{0}^{2}-m_{1}^{2}\right)
^{2}+\epsilon _{1}^{2}\right] }\right] .  \label{annihilation S_0 fermions}
\end{eqnarray}%
The annihilation cross-section into $W$'s is given by:%
\begin{eqnarray}
v_{12}\sigma _{S_{0}S_{0}\rightarrow WW} &=&\frac{\sqrt{m_{0}^{2}-m_{w}^{2}}%
}{16\pi m_{0}^{3}}\Theta \left( m_{0}-m_{w}\right) \left[ 1+\frac{\left(
2m_{0}^{2}-m_{w}^{2}\right) ^{2}}{2m_{w}^{4}}\right]  \notag \\
&&\times \left[ \frac{\left( \lambda _{0}^{\left( 3\right) }\lambda
_{hw}^{(3)}\right) ^{2}}{\left( 4m_{0}^{2}-m_{h}^{2}\right) ^{2}+\epsilon
_{h}^{2}}+\frac{\left( \eta _{01}^{\left( 3\right) }\lambda
_{1w}^{(3)}\right) ^{2}}{\left( 4m_{0}^{2}-m_{1}^{2}\right) ^{2}+\epsilon
_{1}^{2}}\right.  \notag \\
&&+\left. \frac{2\lambda _{0}^{\left( 3\right) }\eta _{01}^{\left( 3\right)
}\lambda _{hw}^{(3)}\lambda _{1w}^{(3)}\left( 4m_{0}^{2}-m_{h}^{2}\right)
\left( 4m_{0}^{2}-m_{1}^{2}\right) }{\left[ \left(
4m_{0}^{2}-m_{h}^{2}\right) ^{2}+\epsilon _{h}^{2}\right] \left[ \left(
4m_{0}^{2}-m_{1}^{2}\right) ^{2}+\epsilon _{1}^{2}\right] }\right] .
\label{annihilation S_0 W}
\end{eqnarray}%
Last, the annihilation cross-section into $Z$'s is:%
\begin{eqnarray}
v_{12}\sigma _{S_{0}S_{0}\rightarrow ZZ} &=&\frac{\sqrt{m_{0}^{2}-m_{z}^{2}}%
}{8\pi m_{0}^{3}}\Theta \left( m_{0}-m_{z}\right) \left[ 1+\frac{\left(
2m_{0}^{2}-m_{z}^{2}\right) ^{2}}{2m_{z}^{4}}\right]  \notag \\
&&\times \left[ \frac{\left( \lambda _{0}^{\left( 3\right) }\lambda
_{hz}^{(3)}\right) ^{2}}{\left( 4m_{0}^{2}-m_{h}^{2}\right) ^{2}+\epsilon
_{h}^{2}}+\frac{\left( \eta _{01}^{\left( 3\right) }\lambda
_{1z}^{(3)}\right) ^{2}}{\left( 4m_{0}^{2}-m_{1}^{2}\right) ^{2}+\epsilon
_{1}^{2}}\right.  \notag \\
&&+\left. \frac{2\lambda _{0}^{\left( 3\right) }\eta _{01}^{\left( 3\right)
}\lambda _{hz}^{(3)}\lambda _{1z}^{(3)}\left( 4m_{0}^{2}-m_{h}^{2}\right)
\left( 4m_{0}^{2}-m_{1}^{2}\right) }{\left[ \left(
4m_{0}^{2}-m_{h}^{2}\right) ^{2}+\epsilon _{h}^{2}\right] \left[ \left(
4m_{0}^{2}-m_{1}^{2}\right) ^{2}+\epsilon _{1}^{2}\right] }\right] .
\label{annihilation S_0 Z}
\end{eqnarray}

The quantities $\epsilon _{h}=m_{h}\Gamma _{h}$ and $\epsilon
_{1}=m_{1}\Gamma _{1}$ are regulators at the respective resonances. The
decay rates $\Gamma _{h}$ and $\Gamma _{1}$ are calculable in perturbation
theory. We have for $h$:%
\begin{eqnarray}
\epsilon _{h\rightarrow f\bar{f}} &=&\frac{\left( \lambda _{hf}\right) ^{2}}{%
8\pi }m_{h}^{2}N_c\left( 1-\frac{4m_{f}^{2}}{m_{h}^{2}}\right) ^{\frac{3}{2}%
}\Theta \left( m_{h}-2m_{f}\right) ;  \notag \\
\epsilon _{h\rightarrow WW} &=&\frac{\left( \lambda _{hw}^{\left( 3\right)
}\right) ^{2}}{8\pi }\left( 1-\frac{4m_{w}^{2}}{m_{h}^{2}}\right) ^{\frac{1}{%
2}}\left[ 1+\frac{\left( m_{h}^{2}-2m_{w}^{2}\right) ^{2}}{8m_{w}^{4}}\right]
\Theta \left( m_{h}-2m_{w}\right) ;  \notag \\
\epsilon _{h\rightarrow ZZ} &=&\frac{\left( \lambda _{hz}^{\left( 3\right)
}\right) ^{2}}{4\pi }\left( 1-\frac{4m_{z}^{2}}{m_{h}^{2}}\right) ^{\frac{1}{%
2}}\left[ 1+\frac{\left( m_{h}^{2}-2m_{z}^{2}\right) ^{2}}{8m_{z}^{4}}\right]
\Theta \left( m_{h}-2m_{z}\right) ;  \notag \\
\epsilon _{h\rightarrow S_{0}S_{0}} &=&\frac{\left( \lambda _{0}^{\left(
3\right) }\right) ^{2}}{32\pi }\left( 1-\frac{4m_{0}^{2}}{m_{h}^{2}}\right)
^{\frac{1}{2}}\Theta \left( m_{h}-2m_{0}\right) ;  \notag \\
\epsilon _{h\rightarrow S_{1}S_{1}} &=&\frac{\left( \lambda _{2}^{\left(
3\right) }\right) ^{2}}{32\pi }\left( 1-\frac{4m_{1}^{2}}{m_{h}^{2}}\right)
^{\frac{1}{2}}\Theta \left( m_{h}-2m_{1}\right) .  \label{decays h}
\end{eqnarray}

For $S_{1}$, we have similar expressions:%
\begin{eqnarray}
\epsilon _{S_{1}\rightarrow f\bar{f}} &=&\frac{\left( \lambda _{1f}\right)
^{2}}{8\pi }m_{1}^{2}N_c\left( 1-\frac{4m_{f}^{2}}{m_{1}^{2}}\right) ^{\frac{3}{%
2}}\Theta \left( m_{1}-2m_{f}\right) ;  \notag \\
\epsilon _{S_{1}\rightarrow WW} &=&\frac{\left( \lambda _{1w}^{\left(
3\right) }\right) ^{2}}{8\pi }\left( 1-\frac{4m_{w}^{2}}{m_{1}^{2}}\right) ^{%
\frac{1}{2}}\left[ 1+\frac{\left( m_{1}^{2}-2m_{w}^{2}\right) ^{2}}{%
8m_{w}^{4}}\right] \Theta \left( m_{1}-2m_{w}\right) ;  \notag \\
\epsilon _{S_{1}\rightarrow ZZ} &=&\frac{\left( \lambda _{1z}^{\left(
3\right) }\right) ^{2}}{4\pi }\left( 1-\frac{4m_{z}^{2}}{m_{1}^{2}}\right) ^{%
\frac{1}{2}}\left[ 1+\frac{\left( m_{1}^{2}-2m_{z}^{2}\right) ^{2}}{%
8m_{z}^{4}}\right] \Theta \left( m_{1}-2m_{z}\right) ;  \notag \\
\epsilon _{S_{1}\rightarrow S_{0}S_{0}} &=&\frac{\left( \eta _{01}^{\left(
3\right) }\right) ^{2}}{32\pi }\left( 1-\frac{4m_{0}^{2}}{m_{1}^{2}}\right)
^{\frac{1}{2}}\Theta \left( m_{1}-2m_{0}\right) ;  \notag \\
\epsilon _{S_{1}\rightarrow hh} &=&\frac{\left( \lambda _{1}^{\left(
3\right) }\right) ^{2}}{32\pi }\left( 1-\frac{4m_{h}^{2}}{m_{1}^{2}}\right)
^{\frac{1}{2}}\Theta \left( m_{1}-2m_{h}\right) .  \label{decays s_1}
\end{eqnarray}
where $N_c$ is equal to $1$  for leptons and $3$ for quarks.


\begin{thebibliography}{99}
\bibitem{Observ} D.~N.~Spergel \textit{et al}. [WMAP Collaboration], Astrophys.~J.~Suppl.~170 (2007) 377 (\texttt{[arXiv:astro-ph/0603449].
});  A.~C.~Pope {\it et al.}  [The SDSS Collaboration], Astrophys.~J.~{\bf 607} (2004) 655 (\texttt{arXiv:astro-ph/0401249}).

\bibitem{dama-libra} R.~Bernabei, P.~Belli and  F.~Cappella \textit{et al}.
[DAMA/LIBRA Collaboration], \texttt{arXiv:1007.0595 [astro-ph.CO]}; Eur.
Phys.~J.~\textbf{C67} (2010) 39 (\texttt{arXiv:1002.1028 [astro-ph.GA]}).

\bibitem{cogent} C.~E.~Aalseth \textit{et al}. [CoGeNT Collaboration],
\texttt{arXiv:1002.4703 [astro-ph.CO]}.

\bibitem{HG} D.~Hooper and L.~Goodenough, \texttt{arXiv:1010.2752 [hep-ph]}.

\bibitem{Groups}  S.~Andreas, T.~Hambye and M.~H.~G.~Tytgat,  JCAP {\bf 0810}, 034 (2008)
   (\texttt{arXiv:0808.0255 [hep-ph]});  S.~Chang, J.~Liu, A.~Pierce, N.~Weiner, I.~Yavin, JCAP {\bf 1008} (2010) 018  (\texttt{arXiv:1004.0697 [hep-ph]}) ;  R.~ Essig, J.~ Kaplan, P.~ Schuster and  N.~ Toro, arXiv:1004.0691 [hep-ph]  ;  S.~Andreas, C.~Arina, T.~Hambye, Fu-Sin Ling, M.~H.~G.~Tytgat, \texttt{arXiv:1003.2595 [hep-ph]} ;  D.~ Hooper, J.~I.~ Collar, J.~ Hall, D.~McKinsey and  C.~Kelso, Phys.~Rev.~\textbf{D88} (2010) 123509 (\texttt{arXiv:1007.1005 [hep-ph].}); R.~Foot,
  Phys.~Lett.~B {\bf 692}, 65 (2010) (\texttt{arXiv:1004.1424 [hep-ph]}); A.~L.~Fitzpatrick, D.~Hooper and K.~M.~Zurek, Phys.~Rev.~\textbf{D81} (2010) 115005 (\texttt{arXiv:1003.0014[hep-ph]}).
.

\bibitem{dave-et-al} R Dav\'{e}, D.~N.~ Spergel, P.~J.~Steinhardt and B.~D.~
Wandelt, Astrophys.~J.~\textbf{547} (2001) 574 (\texttt{astro-ph/0006218}).

\bibitem{mcdonald-2} J.~McDonald, Phys.~Rev.~Lett.~\textbf{88} (2002) 091304 (\texttt{%
hep-ph/0106249}).



\bibitem{xenon100} E.~Aprile \textit{et al}. [XENON100 Collaboration], Phys.~Rev.~Lett.~\textbf{105} (2010) 131302 (\texttt{arXiv:1005.0380 [astro-ph.CO]}); \texttt{arXiv:1005.2615 [astro-ph.CO]}.


\bibitem{collar-mckinsey} J.~I.~Collar and D.~N.~McKinsey, \texttt{%
arXiv:1005.0838 [astro-ph.CO]}.

\bibitem{Ge2010} D.~S.~Akerib {\it et al.}  [CDMS Collaboration],
  arXiv:1010.4290 [astro-ph.CO].
; Z. Ahmed \textit{et al}. [CDMS Collaboration], \texttt{arXiv:1011.2482 [astro-ph]}.


\bibitem{Chann} E.~M.~Drobyshevski,  \texttt{arXiv:0706.3095 [astro-ph]}; R.~Bernabei et al, Eur.~Phys.~J.~\textbf{C53}, 205 (2008); N.~Bozorgnia, G.~B.~Gelmini and  P.~Gondolo, JCAP {\bf 1011} (2010) 019
  (\texttt{arXiv:1006.3110 [astro-ph.CO]}); N.~Bozorgnia, G.~B.~Gelmini and P.~Gondolo,
  JCAP {\bf 1011} (2010) 028  (\texttt{arXiv:1008.3676 [astro-ph.CO]}).


\bibitem{FS}  M.~Fairbairn and T.~Schwetz, JCAP {\bf 0901}, 037 (2009)
  (\texttt{arXiv:0808.0704 [hep-ph]}).

\bibitem{KH} C.~Kelso and D.~Hooper, \texttt{arXiv:1011.3076 [hep-ph]}.

\bibitem{CRESST} W.~Seidel,  WONDER 2010 Workshop, Laboratory Nazionali del Gran Sasso, Italy, March 22-23, 2010 ; IDM 2010 Workshop, Montpellier, France, July
26-30, 2010.

\bibitem{LSP}  J.~Ellis, J.~S.~Hagelin, D.~V.~Nanopoulos, K.~A.~Olive and M.~Srednicki, Nucl.~Phys.~\textbf{B238} (1984) 453.

\bibitem{Kamio}  G.~Jungman, M.~Kamionkowski and K.~Griest,
  Phys.~Rept.~{\bf 267} (1996) 195 (\texttt{arXiv:hep-ph/9506380}).

\bibitem{Nathal} E.~Kuflik, A.~Pierce and K.~M.~Zurek, Phys.~Rev.~\textbf{D81} (2010) 111701;  D.~Feldman, Z.~Liu and P.~Nath,
  Phys.~Rev.~\textbf{D81} (2010) 117701 (\texttt{arXiv:1003.0437 [hep-ph]}).

\bibitem{Betal} A.~Bottino, N.~Fornengo and S.~Scopel, Phys.~Rev.~\textbf{D67} (2003) 063519  (\texttt{arXiv:hep-ph/0212379}); A.~Bottino,
F.~Donato, N.~Fornengo and S.~Scopel, Phys.~Rev.~\textbf{D78} (2008) 083520  (\texttt{arXiv:0806.4099 [hep-ph]}); V.~Niro, A.~Bottino,
N. Fornengo and S. Scopel, Phys.~Rev.~\textbf{D80} (2009) 095019  (\texttt{arXiv:0909.2348 [hep-ph]}); A.~Bottino, F.~Donato, N.~Fornengo
and S.~ Scopel, Phys.~ Rev.~ \textbf{D72} (2005) 083521  (\texttt{arXiv:hep-ph/0508270}).

\bibitem{HT} D.~Hooper and T.~Plehn, Phys.~Lett.~\textbf{B562} (2003) 18  (\texttt{arXiv:hep-ph/0212226}).



\bibitem{silveira-zee} V.~Silveira and A.~Zee, Phys.~Lett.~\textbf{B161}
(1985) 136.

\bibitem{mcdonald-1} J.~McDonald, Phys.~Rev.~\textbf{D50} (1994) 3637.

\bibitem{burgess-pospelov-veldhuis} C.~P.~Burgess, M.~Pospelov and T.~ter
Veldhuis, Nucl.~ Phys.~ \textbf{B619} (2001) 709.

\bibitem{barger et al} V.~Barger, P.~Langacker, M.~McCaskey, M.~J.~Ramsey-Musolf and G.~Shaughnessy, Phys.~Rev.~\textbf{D77} (2008) 035005 (%
\texttt{arXiv:0706.4311 [hep-ph]}).

\bibitem{gonderinger et al} M.~Gonderinger, Y.~Li, H.~Patel and M.~J.~Ramsey-Musolf, JHEP \textbf{053} (2010) 1001, 2010 (\texttt{arXiv:0910.3167
[hep-ph]}).

\bibitem{he-li-li-tandean-tsai} X.G.~He, T.~Li, X.Q.~Li, J.~Tandean and
H.C.~Tsai, Phys.~Rev.~\textbf{D79} (2009) 023521 (\texttt{arXiv:0811.0658
[hep-ph])}.

\bibitem{xenon10-1} J.~Angle \textit{et al}. [XENON Collaboration], Phys.~Rev.~Lett.~\textbf{100} (2008) 021303 (\texttt{arXiv:0706.0039 [astro-ph]}).

\bibitem{cdmsII-1} Z.~Ahmed \textit{et al}. [CDMS Collaboration], Phys.~Rev.~Lett.~\textbf{102} (2009) 011301 (\texttt{arXiv:0802.3530 [astro-ph]});
Z.~Ahmed \textit{et al}. [CDMS Collaboration], Science \textbf{327} (2010) 1619 (\texttt{arXiv:0912.3592 [astro-ph.CO]}).


\bibitem{supercdms} R.~W.~Schnee {\it et al.}  [The SuperCDMS Collaboration],  \texttt{arXiv:astro-ph/0502435}.
.

\bibitem{xenon1t}  E.~Aprile  [Xenon Collaboration],
  J.\ Phys.\ Conf.\ Ser.\  {\bf 203} (2010) 012005.

\bibitem{particle-data-group} K.~Nakamura \textit{et al}. [Particle Data
Group], J.~Phys.~G\textbf{37} (2010) 075021.

\bibitem{about-eta0} The effect of $\eta _{0}$ in the one-real-scalar
extension of the Standard Model is discussed in D.N. Spergel and P. J.
Steinhardt, Phys.~Rev.~Lett.~\textbf{84} (2000) 3760 (\texttt{%
astro-ph/9909386}). See also \cite{mcdonald-2}.

\bibitem{Kolb-Turner} E.~W.~Kolb and M.~S.~Turner, The Early Universe, Addison-Wesley, (1998).

\bibitem{WMAP2010} E.~Komatsu ~{\it et al.},  \texttt{arXiv:1001.4538 [astro-ph.CO]}.

\bibitem{Weinberg} S.~Weinberg, ``Cosmology'',  Oxford University Press, (2008).


\bibitem{abada-nasri-1} A.~Abada and S.~Nasri, work in progress.

\bibitem{exp-data} R.~Gaitskell, V.~ Mandic, and J.~Filippini, SUSY Dark Matter/Interactive Direct Detection Limit
Plotter, http://dmtools.berkeley.edu/limitplots.
\end{thebibliography}
\end{document}